\documentclass[review,3p,number]{elsarticle}
\usepackage{eurosym}
\usepackage{amsfonts}
\usepackage{graphicx}
\usepackage{color}
\usepackage{amsmath}
\usepackage{amssymb}
\usepackage{latexsym}

\def\epsilon{\varepsilon}

\makeatletter

\@addtoreset{equation}{section}
\makeatother
\biboptions{sort,compress}

\begin{document}

\begin{frontmatter}
\title{Families of solitons in Bragg supergratings}

\cortext[cor1]{Corresponding author}

\author{Boris A. Malomed}
\address{Department of Physical Electronics, School of Electrical Engineering,
Faculty of Engineering, Tel Aviv University, Tel Aviv 69978, Israel
}
\ead{malomed@post.tau.ac.il}

\author{Thomas Wagenknecht}
\address{Department of Applied Mathematics, University of Leeds, \\ Leeds LS2 9JT, UK}
\ead{thomas@maths.leeds.ac.uk}

\author{Kazuyuki Yagasaki\corref{cor1}}
\address{Mathematics Division, Department of Information Engineering,\\
 Niigata Unversity, Niigata 950-2181, Japan}
\ead{yagasaki@ie.niigata-u.ac.jp}

\begin{abstract}
We study fundamental optical gap solitons in the model of a fiber Bragg grating (BG), which is
subjected to a periodic modulation of the local reflectivity, giving rise to a supergrating. In
addition, the local refractive index is also periodically modulated with the same period.
It is known that the supergrating opens an infinite system of new bandgaps in the BG's spectrum.
We use a combination of analytical and
computational methods to show that each emerging bandgap is filled with gap
solitons (GSs), including asymmetric ones and bound states of the GSs. In particular, bifurcations of the GSs
created by the supergrating are studied in terms of a geometric analysis.
\end{abstract}


\end{frontmatter}

\section{Introduction}

Bragg gratings (BGs) are light-controlling structures produced by a periodic
variation of the refractive index along an optical fiber or waveguide.
Devices based on fiber gratings, such as dispersion compensators, sensors,
elements of laser cavities, etc., are widely used in optical systems \cite%
{Ka:99}. Gap solitons (GSs) in fiber gratings, alias BG solitons \cite%
{StSi:94}, are supported through the balance between the BG-induced
dispersion, which incorporates the bandgap in the system's linear spectrum,
and the Kerr nonlinearity of the fiber or waveguide. Analytical solutions
for BG solitons in the standard fiber-grating model are well known \cite%
{AcWa:89,ChJo:89}. Slightly more than half of the analytical found family is
stable, as first demonstrated, within the framework of the variational
approximation, in Ref. \cite{MaTa:94}, and then in a consistent numerical
form in Ref. \cite{BaPe:98}. Later, an accurate numerical technique for
simulations of GSs was elaborated in Ref. \cite{Derks}, and a rigorous
analysis of their stability was developed in Ref. \cite{Chugunova}

Following the theoretical prediction, BG solitons were created in the
experiment, using a relatively short BG written in the cladding of an
optical fiber \cite{EgSl:96} and a virtual optically-induced grating \cite%
{Coen:04}. In particular, an essential physical achievement was the creation
of slow BG solitons moving at the velocity much lower than the velocity of
light $c$ in vacuum (namely, $0.16c$ \cite{Mok:06}).

GSs are one of fundamental species of solitons in optics, as well as in
other nonlinear media. In particular, the GS concept was extended to
Bose-Einstein condensates (BECs), where such solitons were predicted in a
condensate loaded into a periodic potential (optical lattice) \cite%
{BaKo:02,OsKi:03,Br:04}, and then created in the experiment \cite{EiAn:04}.

An issue of great significance to the fundamental studies and applications
is the development of methods for control of BG solitons. One of them is
\textit{apodization} \cite{EgSt:99,SlEg:98,MaMa:04,Mok:06}, i.e., the use of
a grating with the Bragg reflectivity gradually varying along the fiber (or
waveguide). In particular, a possibility was predicted to slow down the
soliton and eventually bring it to a halt in a properly apodized BG \cite%
{MaMa:04}. Experimentally, it has been shown that the apodization helps to
couple solitons into fiber gratings \cite{EgSt:99}, and may also split them
\cite{SlEg:98}. The retardation of the BG solitons to the above-mentioned
low velocity, $0.16c$, was also performed with the help of the apodization
\cite{Mok:06}.

Technologies which make it possible to fabricate regular fiber gratings with
periodic apodization, i.e., an effective \textit{superlattice} built on top
of the BG, are well established \cite{Ru:86}. Analysis of the light
propagation in the so developed \textit{supergratings} was 
performed in \cite{BrSt:97}. It was shown that the supergrating
gives rise to extra gaps in the system's spectrum (``Rowland ghost
gaps"). Solitons in the gaps were looked for in an approximation
based on ``supercoupled-mode equations", which, essentially, assume
that a soliton amounts to a slowly varying envelope of the
supergrating's Bloch function(s). As is known, such an approximation
(the averaging method) generally applies to the description of GSs
near bandgap edges \cite{Br:04}. A related problem was considered in
\cite{LoOs:05,PoKe:05}, \textit{viz}., GSs in the BEC loaded into a
double-periodic optical lattice. The period doubling opens an
additional narrow ``mini-gap", where stable GSs exist.

The objective of our work is to investigate the existence of GSs in harmonic
superlattices created on top of the usual BG in fibers or waveguides with
the intrinsic Kerr nonlinearity, extending the analysis initiated in \cite%
{YaMe:05}. Our model has periodic potentials in one spatial dimension.
Analytical approaches for GSs in two or more spatial dimensions were also
developed recently in \cite{PeSc:07,DoPeSc:09,DoUe:09,IlWe:10}.

The paper is
organized as follows. We first give our mathematical model of the
supergrating and analyze the bandgap structure in the model, using its
dispersion relation, in Section~\ref{s:prl}. This constitutes a non-trivial
extension of the classical spectral theory for the Mathieu equation \cite%
{SmJo:99}, and predicts parameter values at which GSs may be expected to
exist in the model. The sequence of new bandgaps not present in the
unmodulated BG is produced. 

In Sections~\ref{s:Mel} and \ref{s:av} we establish the existence of GSs in
the newly opened bandgaps, using a combination of analytical and numerical
techniques. We find various families of GSs in the supergrating, including
those which do not exist in the ordinary GS model (in particular, stable
asymmetric solitons, and multi-soliton bound states). Using the Melnikov
method, we first prove that GSs in the original BG model have counterparts
in the extended model. A similar treatment was used for a perturbed
Hamiltonian partial differential equations in \cite{Ka:01}. Then, the method
of averaging is used to demonstrate the emergence of GSs in the new
bandgaps. The most important result is that in each new bandgap, there is a
single family of fundamental symmetric GSs which \emph{entirely} fills the
gap. Such GSs near band edges in nonlinear Schr\"{o}dinger equations with
periodic potentials were also studied by different approaches in \cite%
{HwAkYa:11,PeSuKi:04}. In Section~\ref{s:bif} we discuss bifurcations of GSs
from a geometric point of view and illustrate this general approach in
numerical computations for our problem.

The majority of the newly established GSs are \emph{stable} solutions of the
model equations. In particular, the family of fundamental GSs in the new
bandgaps is stable even for negative frequencies, when the ordinary GSs are
unstable \cite{MaTa:94,BaPe:98}. We do not study the stability of GSs in
this paper in detail, referring to \cite{YaMe:05} for the verification of
the stability by means of direct simulations.

\section{Preliminaries}

\label{s:prl}

\subsection{The supergrating equations}

The standard model of the Bragg grating (BG) in an optical fiber is based on
the equations \cite{ChJo:89}
\begin{equation}  \label{e:standard}
\begin{split}
&
iu_{t}+iu_{x}+v+\left(|v|^{2}+\textstyle\frac{1}{2}|u|^{2}\right) u = 0,
\\
&
iv_{t}-iv_{x}+u+\left(|u|^{2}+\textstyle\frac{1}{2}|v|^{2}\right) v = 0,%
\end{split}%
\end{equation}
where $u$ and $v$ are amplitudes of the right- and left-propagating waves in
the fiber. In these equations, $x$ is the coordinate along the fiber, and $t$
is time. A family of exact soliton solutions of (\ref{e:standard}) have been
found in \cite{AcWa:89, ChJo:89}.

Here we will consider a situation where the BG is subjected to periodic
modulation in $x$, with a period $2\pi /k$, which gives rise to a \textit{%
supergrating}. The general form of the extended model reads
\begin{equation}
\begin{split}
&
iu_{t}+iu_{x}+\left[ 1-\varepsilon \cos \left( kx\right) \right] v+\mu \cos
\left( kx+\delta \right) ~u+\left( |v|^{2}+\textstyle\frac{1}{2}%
|u|^{2}\right) u = 0, \\
&
iv_{t}-iv_{x}+\left[ 1-\varepsilon \cos \left( kx\right) \right] u+\mu \cos
\left( kx+\delta \right) ~v+\left( |u|^{2}+\textstyle\frac{1}{2}%
|v|^{2}\right) v = 0.%
\end{split}
\label{e:uv}
\end{equation}

The real parameter $\varepsilon >0$ accounts for the periodic modulation of
the BG strength. In real fiber gratings, it may be typically $1$ cm \cite%
{Ru:86,BrSt:97}, while the total length of the grating may be up to $1$ m.
The modulation is implemented through the change imposed on the local
variation of the refractive index in the fiber's cladding and may also
affect the local index felt by each wave. This is accounted for by the
perturbation parameter $\mu \geq 0$ and a phase shift $\delta $. In
practice, the harmonic superlattice corresponding to Eqs. (\ref{e:uv}) may
be implemented if the optical interference pattern, which is used to burn
the grating into the fiber, is created with spatial beatings. We consider
stationary GSs, which are found via the ansatz
\begin{equation*}
u(x,t)=\exp (-i\omega t)U(x),\quad v(x,t)=\exp (-i\omega t)V(x).
\end{equation*}%
Substituting this into Eqs. (\ref{e:uv}), we obtain
\begin{equation}
\begin{split}
&
\omega U+iU^{\prime }+\left[ 1-\varepsilon \cos (kx)\right] V+\mu \cos
\left( kx+\delta \right) U+\left( |V|^{2}+\textstyle\frac{1}{2}%
|U|^{2}\right) U = 0, \\
&
\omega V-iV^{\prime }+\left[ 1-\varepsilon \cos (kx)\right] U+\mu \cos
\left( kx+\delta \right) V+\left( |U|^{2}+\textstyle\frac{1}{2}%
|V|^{2}\right) V = 0,%
\end{split}
\label{e:uvans}
\end{equation}%
where the prime stands for $d/dx$.

\subsection{Linear analysis}

\label{s:linear}

We first analyze the bandgap structure of (\ref{e:uv}) by looking for
solutions of the linearized system in the form of
\[
u(x,t)=\exp \left( iqx-i\omega t\right) U(x),\quad v(x,t)=\exp \left(
iqx-i\omega t\right) V(x),
\]
with $q$ being the propagation constant. Thus, we have to find periodic
solutions of the linearization of (\ref{e:uvans})
\begin{equation}
\begin{split}
&
(\omega -q)U+iU^{\prime }+\left[ 1-\varepsilon \cos (kx)\right] V+\mu \cos
\left( kx+\delta \right) U = 0, \\
&
(\omega +q)V-iV^{\prime }+\left[ 1-\varepsilon \cos (kx)\right] U+\mu \cos
\left( kx+\delta \right) V = 0.%
\end{split}
\label{e:uvlin}
\end{equation}%
Note that in the unperturbed problem, with $\varepsilon ,\mu =0$, the
dispersion relation is given by $\omega ^{2}=q^{2}+1$, so that we find the
well-known gap in the spectrum $-1<\omega <1$.

Let us consider the case of $\mu=0$.
For $\varepsilon>0$ non-trivial solutions of (\ref{e:uvlin}) emerge due to
parametric resonance, similar to the situation in the classical Mathieu
equation \cite{SmJo:99}. Consequently, we expect new gaps to open up at
points $\omega_{\pm m}= \pm \sqrt{1 + (mk)^2/ 4}$, $m= \pm 1, \pm 2, \ldots$%
. We will denote the corresponding gaps by $\mathbf{1^\pm}, \mathbf{2^\pm},
\ldots $, while the central gap will be denoted by $\mathbf{0}$. In the
central gap $\mathbf{0}$, GSs for $\varepsilon,\mu\neq 0$ are found as
perturbations of the soliton for $\varepsilon,\mu=0$, established in \cite%
{AcWa:89}. We will use Melnikov's method \cite{GuHo:83} to prove the
existence of gap solitons in Section~\ref{s:Mel}. For the non-central gaps,
however, we will apply the averaging method to investigate the emergence of
(small) GSs for $\varepsilon,\mu\neq 0$ in Section~\ref{s:av}.

Straightforward perturbation theory for parametrically excited systems
yields approximation to the gaps for small $\varepsilon >0$. For example, an
approximation to the gap with $\omega $ close to $\omega _{\pm 1}$, i.e.
close to the unperturbed first gap at $\varepsilon =0$, can be obtained by
looking for solutions of (\ref{e:uvlin}) to lowest
order approximation
\begin{align*}
U(x)=& A+\alpha _{1}\cos (kx)+\alpha _{2}\sin (kx),\\
V(x)=& A+\beta _{1}\cos (kx)+\beta _{2}\sin (kx).%
\end{align*}
Using this ansatz we derive at the solvability condition at order $%
O(\varepsilon ^{2})$,
\begin{equation}
\begin{array}{c}
(\omega ^{2}-q^{2}-1)\left[ (\omega ^{2}-q^{2}-1-k^{2})^{2}-4k^{2}q^{2}%
\right]  \\
\qquad \qquad \qquad =\varepsilon ^{2}\left[ (\omega
^{2}-q^{2})^{2}-k^{2}(\omega ^{2}+q^{2}+1)-1\right] .%
\end{array}
\label{e:solv}
\end{equation}%
Observe that $\omega ^{2}\approx q^{2}+1$, since we are interested in the
gap close to the unperturbed one with $\varepsilon =0$. Moreover, note that
at the boundaries of the gaps periodic solutions with wavenumber $k$ or $k/2$
exist \cite{SmJo:99}. Hence, being interested in solutions with $q\approx k/2
$ we introduce
\[
P:=\omega ^{2}-q^{2}-1,\quad Q:=k^{2}-4q^{2},
\]
and expand (\ref{e:solv}) for small $P,Q$ to obtain
\[
2k^{2}P^{2}+((k^{2}-2)\varepsilon ^{2}-k^{2}Q)P+2k^{2}(q^{2}+1)\varepsilon
^{2}=0.
\]
A new gap appears if this quadratic equation for $P$ has no real solutions.
Thus we obtain as the condition for the gap, that
\[
\left\vert Q-\frac{k^{2}-2}{k^{2}}\varepsilon ^{2}\right\vert <2\varepsilon
\sqrt{k^{2}+4}.
\]
In terms of the original parameter $\omega $ we find that the gaps open up
at $\omega _{\pm 1}=\sqrt{1+k^{2}/4}$ and their widths are
\[
\Delta \omega =\frac{q\Delta q}{\omega }=\frac{k}{2}\varepsilon.
\]
Similar expressions can be obtained for the higher-order gaps $\mathbf{%
m^{\pm }}$ for $m\geq 2$ showing that the gap width of the band
corresponding to $m$ scales with $\varepsilon ^{|m|}$.
We can treat the case of $\mu\neq 0$ similarly.

These perturbative results can be extended numerically by detecting
parameter values, for which Eqs.~(\ref{e:uvlin}) possesses periodic
solutions with wavenumber $k$ or $k/2$. It is well known that these
parameters form the boundary of the gaps in the spectrum (compare with
above).

\begin{figure}[t]
\begin{center}
\includegraphics[width=1.0 \textwidth]{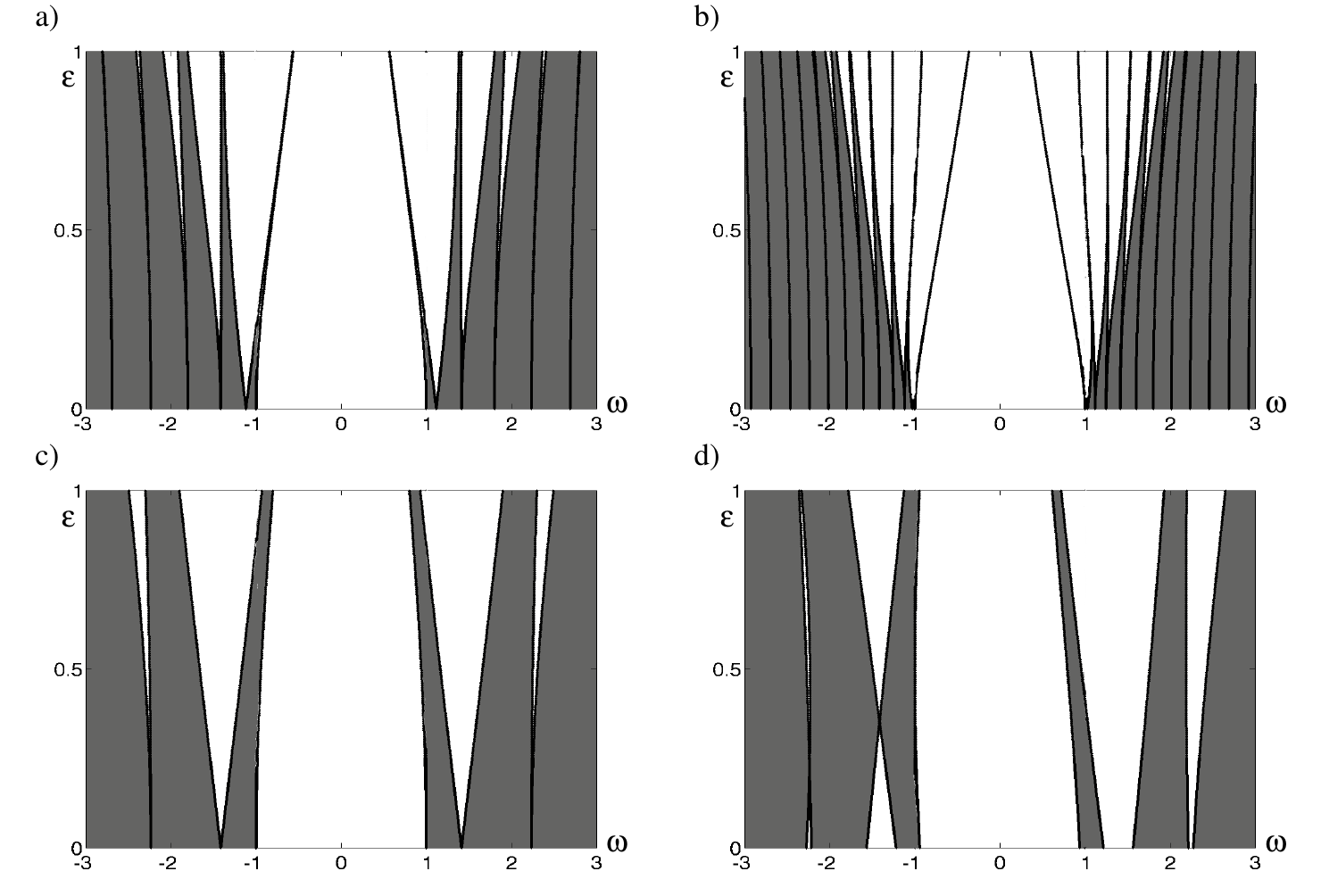}
\caption{The linear spectrum of Eq. (\protect\ref{e:uvlin}) in the $(\protect%
\omega ,\protect\varepsilon )$ plane. Shaded and white areas are,
respectively, bands and gaps. The diagrams show the spectrum for a) $k=1$, $%
\protect\mu=0$, b) $k=0.5$, $\protect\mu=0$, c) $k=2$, $\protect\mu=0$, and
d) $k=2$, $\protect\mu=0.5$, $\protect\delta=0$.}
\label{f:linear}
\end{center}
\end{figure}

Following \cite{Tr:00} we have computed these parameter values as solutions
of the corresponding eigenvalue problem in Matlab. The results for several
combinations of parameter values $k$ and $\mu$ are shown in Fig.~\ref%
{f:linear}. In this figure the shaded and white areas stand for bands and
gaps, respectively. It is found that varying $k$ merely changes the width of
bands and gaps with respect to $\omega$, whereas changing $\mu$ leads to
more complex effects. In particular, for $\mu \neq 0$ gaps opening at $%
\varepsilon,\mu=0$ can vanish again when $\varepsilon$ or $\mu$ are
increased.

\section{Existence of gap solitons in the central gap}

\label{s:Mel}

In this and next sections we investigate the existence of GSs in Eq.~(\ref%
{e:uv}), using both analytical and numerical techniques. These solitons are
described by homoclinic solutions to the origin $(U,V)=(0,0)$ in Eqs.~(\ref%
{e:uvans}).


Before we present the computations let us take a closer look at properties
of Eqs.~(\ref{e:uvans}). It will be important that these equations allow for
a symmetry reduction by setting $V = -U^*$, where '$*$' denotes complex
conjugation. In the following, we will thus only consider the reduced
equation
\begin{equation}  \label{e:u}
i U^{\prime}+\omega U-(1-\varepsilon\cos kx)U^{\ast}+\mu\cos(kx+\delta)U +%
\frac{3}{2}|U|^{2}U=0.
\end{equation}
Let $U=a+ib$, where $a,b\in\mathbb{R}$. Eq.~(\ref{e:u}) is rewritten as
\begin{equation}
\begin{split}
a^{\prime}=&-[\omega+1-\varepsilon\cos kx+\mu\cos(kx+\delta)]b -\frac{3}{2}%
(a^{2}+b^{2})b, \\
b^{\prime}=&[\omega-1+\varepsilon\cos kx+\mu\cos(kx+\delta)]a +\frac{3}{2}%
(a^{2}+b^{2})a.
\end{split}
\label{e:ab}
\end{equation}
We will consider \eqref{e:ab} as a dynamical system in the three-dimensional
phase space $\mathbb{R}^2\times\mathbb{S}^1$, with $\mathbb{S}^1$ as the
circle of length $2\pi/k$.


Note that if $\mu\sin\delta=0$, then Eqs.~(\ref{e:ab}) are \textit{reversible%
}, that is, invariant under compositions of time reversal and the (linear)
involutions,
\begin{align*}
& R_1:(a,b,x)\mapsto(a,-b,-x),\qquad R_1^{\prime}:\left(a,b,x-\frac{\pi}{k}%
\right) \mapsto\left(a,-b,\frac{\pi}{k}-x\right), \\
& R_2:(a,b,x)\mapsto(-a,b,-x),\qquad R_2^{\prime}:\left(a,b,x-\frac{\pi}{k}%
\right) \mapsto\left(-a,b,\frac{\pi}{k}-x\right).
\end{align*}
We refer to \cite{LaRo:98} for a general review of reversible systems. Of
particular importance for us will be \textit{symmetric} homoclinic orbits,
which are mapped to itself under the action of $R_{1,2}$ or $%
R_{1,2}^{\prime} $. It is well known that an orbit is symmetric if and only
if it intersects the invariant plane of the involution.

Now we assume that $0\le \varepsilon,\mu\ll1$ and $-1<\omega<1$, and set $%
\omega=\cos\theta$ for some $\theta\in(0,\pi)$. The cases of $|\omega|>1$
and $|\omega|\approx 1$ will be treated in Section~\ref{s:av} and \ref%
{a:bound}, respectively.

For $\varepsilon,\mu=0$ Eqs.~(\ref{e:ab}) become a planar Hamiltonian
system,
\begin{equation}
\begin{split}
a^{\prime}=&-(\omega+1)b-\frac{3}{2}(a^{2}+b^{2})b, \\
b^{\prime}=&(\omega-1)a+\frac{3}{2}(a^{2}+b^{2})a,
\end{split}
\label{e:ab0}
\end{equation}
with a Hamiltonian
\[
H(a,b)=\frac{1}{2}[(\omega-1)a^2+(\omega+1)b^2]+\frac{3}{8}(a^2+b^2)^2,
\]
where $a$ and $b$ represent the canonical momentum and coordinates,
respectively. The equilibrium at the origin is a hyperbolic saddle in (\ref%
{e:ab0}) and has a pair of homoclinic orbits
\begin{align}
(a_\pm(x),b_\pm(x))=&\left( \pm2\sqrt{\frac{2}{3}}\sin\theta\cos\!\left(%
\frac{\theta}{2}\right) \frac{\cosh(x\sin\theta)}{\cosh(2x\sin\theta)+\cos%
\theta},\right.  \notag \\
& \qquad \left. \pm2\sqrt{\frac{2}{3}}\sin\theta\sin\!\left(\frac{\theta}{2}%
\right) \frac{\sinh(x\sin\theta)}{\cosh(2x\sin\theta)+\cos\theta} \right),
\label{e:homo0}
\end{align}
which are symmetric under the involution $R_1$. When $\varepsilon$ and/or $%
\mu$ are nonzero but sufficiently small, the origin is still a hyperbolic
saddle in \eqref{e:ab} and has two-dimensional stable and unstable manifolds
which may intersect transversely. Such intersection yields transverse
homoclinic orbits to the saddle at the origin, which persist under variation
of the parameters \cite{Wi:90}. Along certain curves in the parameter space,
however, the intersection of the manifolds may become tangential, resulting
in bifurcations of the homoclinic orbits. See Section \ref{s:bif} for
details.

Here we aim to prove the existence of such transverse homoclinic orbits in
Eqs. (\ref{e:ab}) by means of the Melnikov method \cite{GuHo:83}. {From
formula (4.5.6) in \cite{GuHo:83}, we derive the Melnikov functions $M_{\pm
}(x_{0})$ for $(a_{\pm }(x),b_{\pm }(x))$ as}
\begin{align}
M_{\pm }(x_{0})=& \varepsilon \int_{-\infty }^{\infty }[2\omega +3(a_{\pm
}^{2}(x)+b_{\pm }^{2}(x))]\,a_{\pm }(x)\,b_{\pm }(x)\cos k(x+x_{0})dx  \notag
\\
& +2\mu \int_{-\infty }^{\infty }a_{\pm }(x)\,b_{\pm }(x)\cos
[k(x+x_{0})+\delta ]dx  \notag \\
=& -\varepsilon \left( \int_{-\infty }^{\infty }[2\omega +3(a_{\pm
}^{2}(x)+b_{\pm }^{2}(x))]\,a_{\pm }(x)\,b_{\pm }(x)\sin kx\,dx\right) \sin
kx_{0}  \notag \\
& -2\mu \left( \int_{-\infty }^{\infty }a_{\pm }(x)\,b_{\pm }(x)\sin
kx\,dx\right) \sin (kx_{0}+\delta ),  \label{e:Mel}
\end{align}%
{where we used the fact that $a_{\pm }(x)$ and $b_{\pm }(x)$ are even and
odd functions of $x$, respectively.} Substituting (\ref{e:homo0}) into (\ref%
{e:Mel}) and using the method of residues, we compute
\begin{align}
& \int_{-\infty }^{\infty }a_{\pm }(x)\,b_{\pm }(x)\sin kx\,dx  \notag \\
& =\frac{2}{3}\sin ^{3}\theta \int_{-\infty }^{\infty }\frac{\sinh (2x\sin
\theta )}{[\cosh (2x\sin \theta )+\cos \theta ]^{2}}\sin kx\,dx  \notag \\
& =\frac{\pi k}{3}\,\mathrm{cosech}\!\left( \frac{k\pi }{2\sin \theta }%
\right) \sinh \!\left( \frac{k\theta }{2\sin \theta }\right)  \label{e:int1}
\end{align}%
and
\begin{align}
& \int_{-\infty }^{\infty }[a_{\pm }^{2}(x)+b_{\pm }^{2}(x)]a_{\pm
}(x)\,b_{\pm }(x)\sin kx\,dx  \notag \\
& =\frac{8}{9}\sin ^{5}\theta \int_{-\infty }^{\infty }\frac{\sinh (2x\sin
\theta )}{[\cosh (2x\sin \theta )+\cos \theta ]^{3}}\sin kx\,dx  \notag \\
& =\frac{\pi k}{9}\,\mathrm{cosech}\!\left( \frac{k\pi }{2\sin \theta }%
\right) \left[ -2\sinh \!\left( \frac{k\theta }{2\sin \theta }\right) \cos
\theta +k\cosh \!\left( \frac{k\theta }{2\sin \theta }\right) \right] .
\label{e:int2}
\end{align}%
See \ref{a:cal} for derivations of (\ref{e:int1}) and (\ref{e:int2}). Hence,
the Melnikov functions become
\begin{align*}
M_{\pm }(x_{0})=& -\frac{\pi k^{2}}{3}\,\varepsilon \,\mathrm{cosech}%
\!\left( \frac{k\pi }{2\sin \theta }\right) \cosh \!\left( \frac{k\theta }{%
2\sin \theta }\right) \sin kx_{0}  \notag \\
& -\frac{2\pi k}{3}\,\mu \,\mathrm{cosech}\!\left( \frac{k\pi }{2\sin \theta
}\right) \sinh \!\left( \frac{k\theta }{2\sin \theta }\right) \sin
(kx_{0}+\delta ),
\end{align*}%
where we used the relation $\omega =\cos \theta $.

Let us assume that
\begin{equation}
\delta\neq\pi\quad\mbox{ or }\quad \varepsilon \neq\frac{2\mu}{k}%
\tanh\!\left(\frac{k\theta}{2\sin\theta}\right).  \label{e:con}
\end{equation}
Then we easily see that $M_\pm(x_0)$ has simple zeros at
\begin{equation}
x_0=\bar{x}_0,\ \bar{x}_0+\frac{\pi}{k}\mod \frac{2\pi}{k},  \label{e:x0}
\end{equation}
where
\[
\bar{x}_0 =-\frac{1}{k}\arctan\left( \frac{2\mu\sin\delta} {%
2\mu\cos\delta+k\varepsilon\coth(k\theta/2\sin\theta)} \right).
\]
This implies that there exist transverse homoclinic orbits near $%
(a,b)=(a_\pm(x-\bar{x}_0),b_\pm(x-\bar{x}_0))$ and $(a_\pm(x-\bar{x}%
_0-\pi/k),b_\pm(x-\bar{x}_0-\pi/k))$. In particular, when $\mu\sin\delta=0$,
we have $\bar{x}_0=0$ so that the first and second orbit are symmetric under
$R_1$ and $R_2$, respectively. Thus, we can prove that there exist GSs with $%
V=-U^*$ in (\ref{e:uv}).

\begin{figure}[t]
\begin{center}
\includegraphics[scale=0.65]{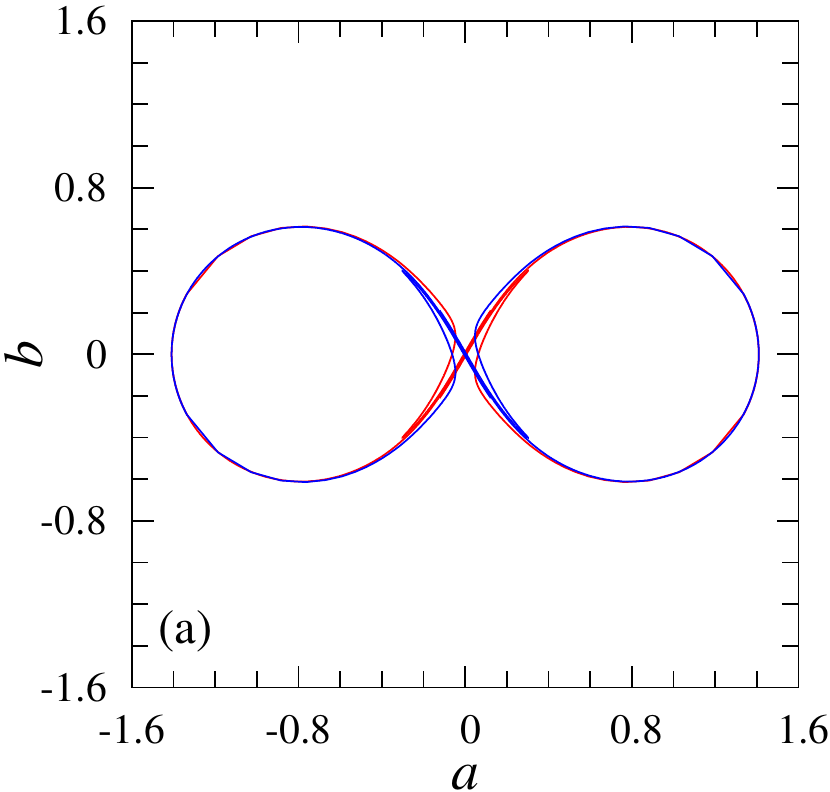}\qquad %
\includegraphics[scale=0.65]{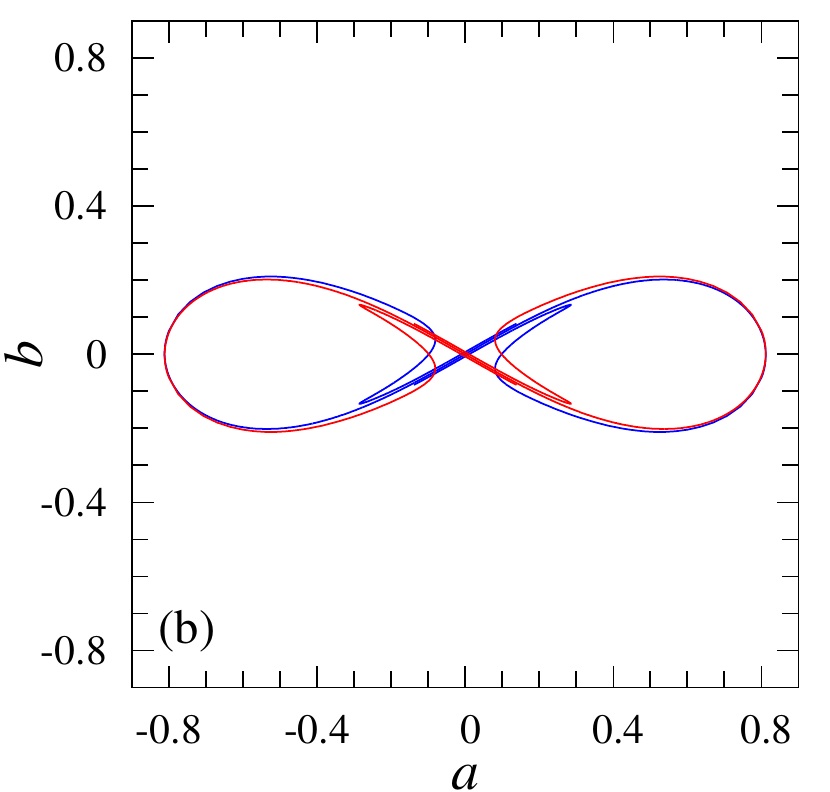}
\caption{Stable manifold (red) and unstable manifold (blue) of the origin in
the Poincar\'{e} section $x=0\mod 2\protect\pi$ for $\protect\varepsilon%
=0.01 $, $\protect\mu=0$ and $k=1$: (a) $\protect\omega=-0.5$; (b) $\protect%
\omega=0.5$. }
\label{f:im0}
\end{center}
\end{figure}

\setlength{\unitlength}{1mm}
\begin{figure}[t]
\begin{center}
\begin{minipage}{.4\textwidth}
\begin{center}
\includegraphics[width=\textwidth]{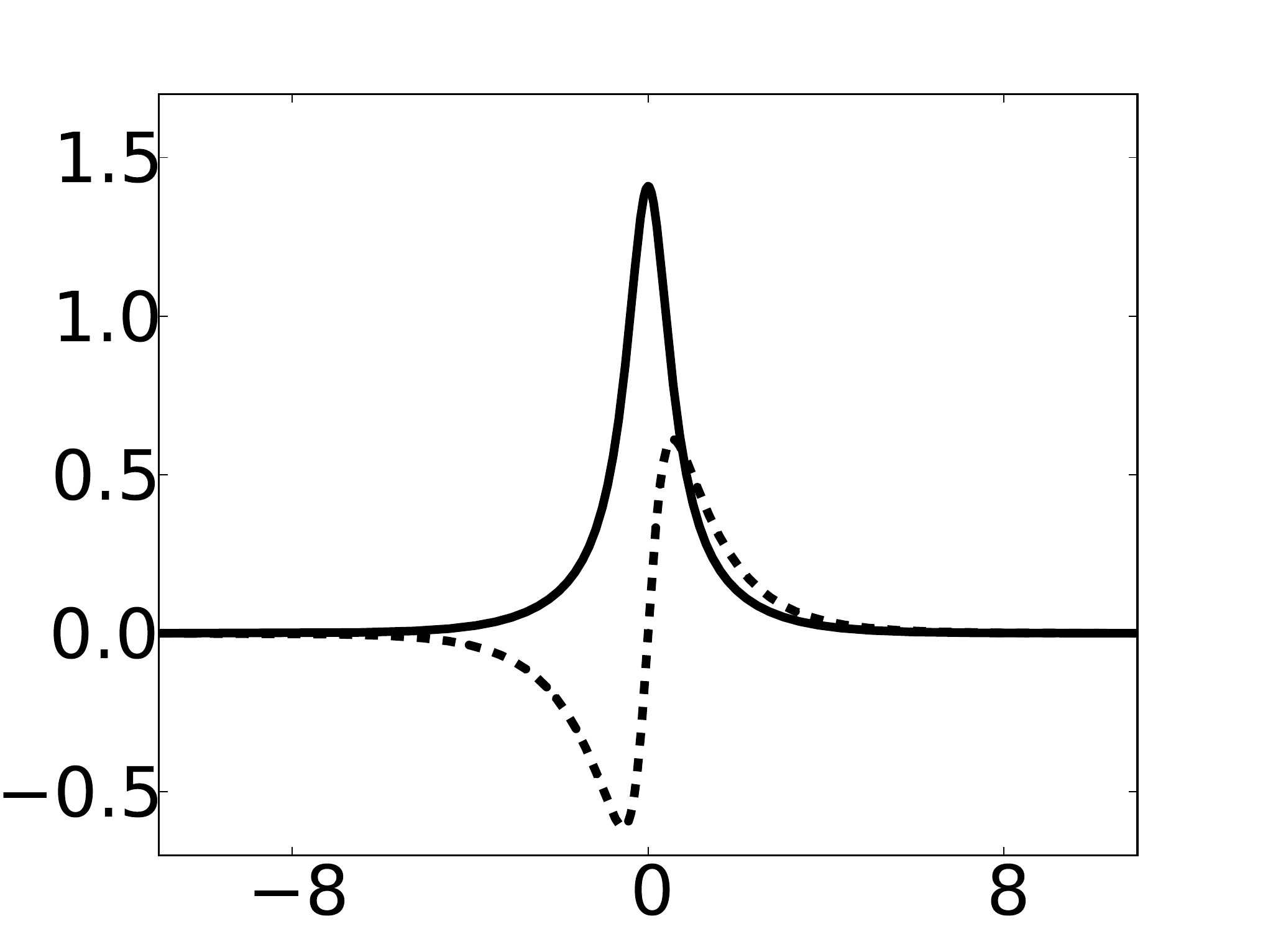}
(a)  $\omega=-0.5$
\end{center}
\end{minipage}
\qquad
\begin{minipage}{.4\textwidth}
\begin{center}
\includegraphics[width=\textwidth]{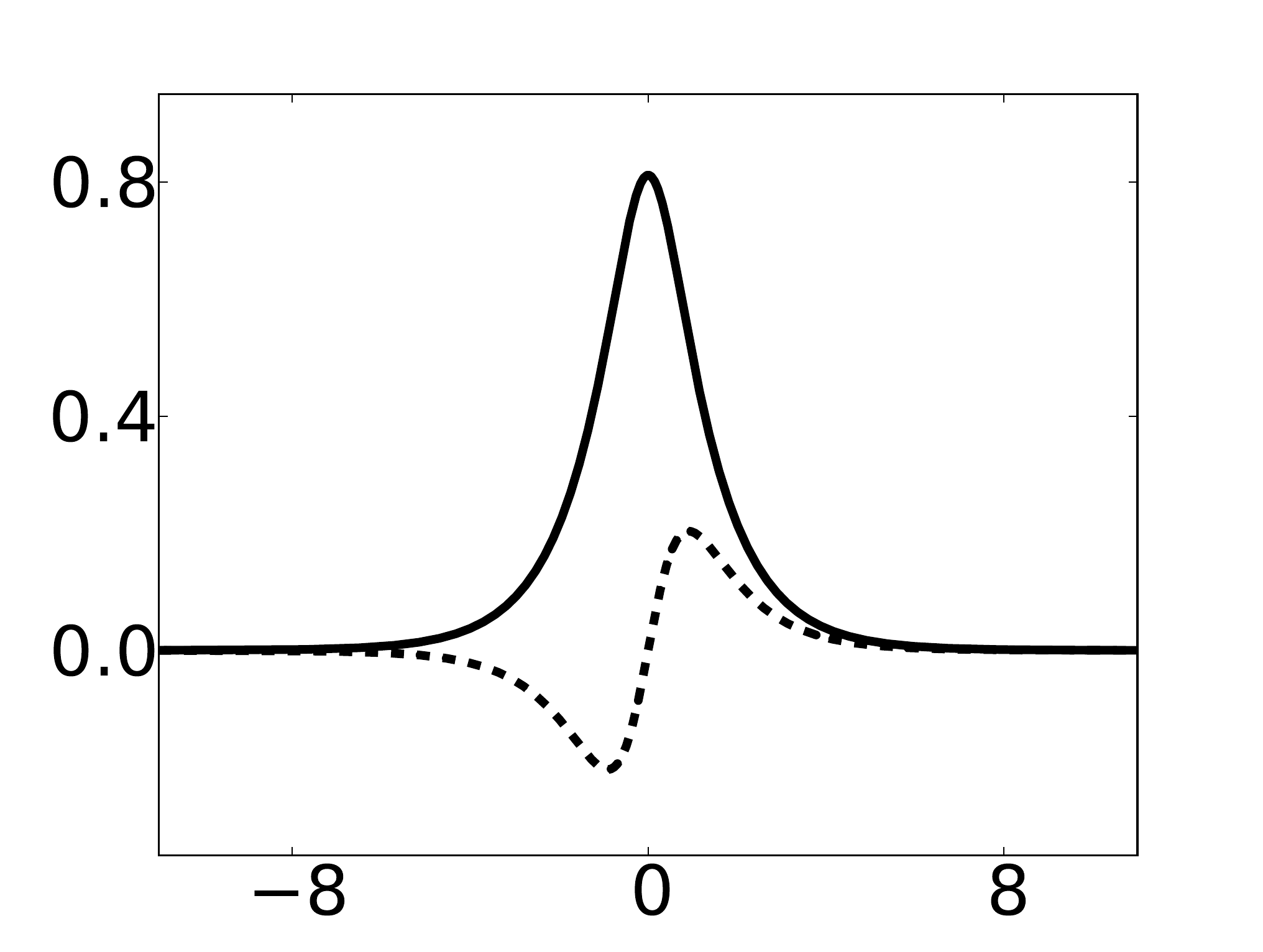}
(b)  $\omega=0.5$
\end{center}
\end{minipage}
\caption{Symmetric GSs with $V=-U^*$ of (\protect\ref{e:uv}) in the central
gap with $k=1$, $\protect\varepsilon=0.01$ and $\protect\mu=0$. The solid
curve shows the real part and the dashed curve the imaginary part of the
solution. }
\label{f:big0}
\end{center}
\end{figure}

\begin{figure}[tbp]
\begin{center}
\begin{minipage}{.4\textwidth}
\begin{center}
\includegraphics[width=\textwidth]{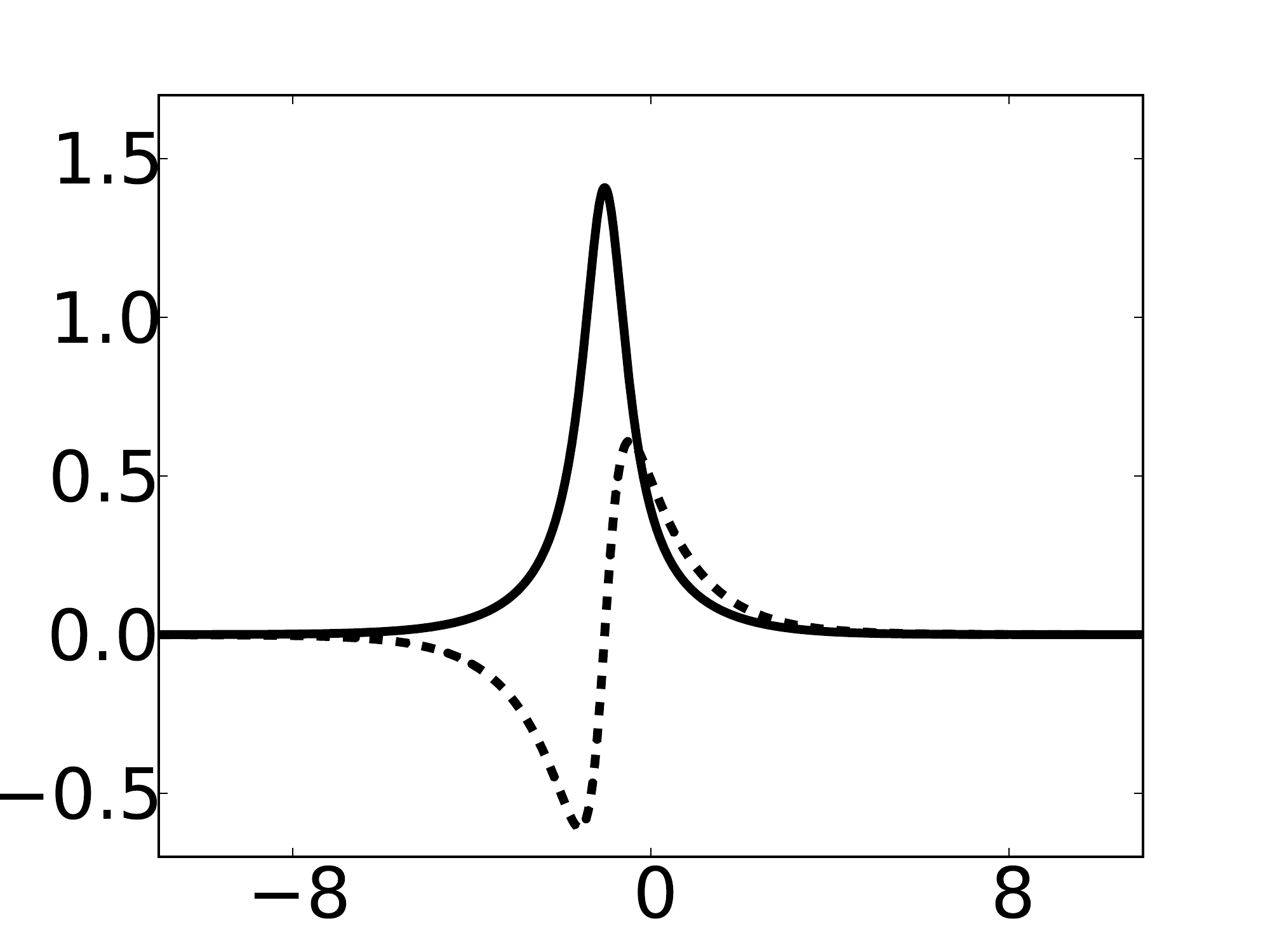}\\
(a) $\omega=-0.5$
\end{center}
\end{minipage}
\qquad
\begin{minipage}{.4\textwidth}
\begin{center}
\includegraphics[width=\textwidth]{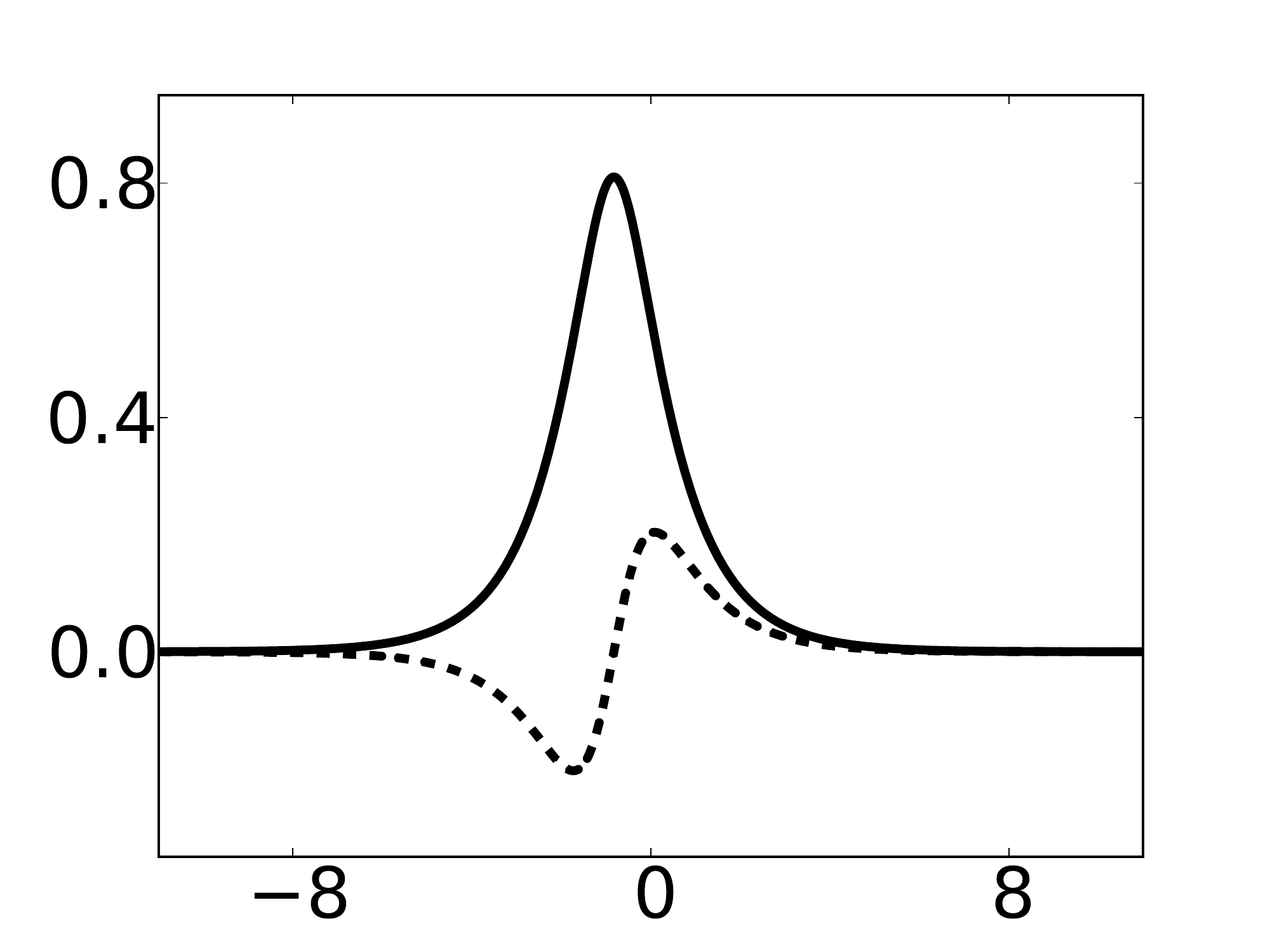}\\
(b) $\omega=0.5$
\end{center}
\end{minipage}
\caption{Asymmetric GSs with $V=-U^*$ of (\protect\ref{e:uv}) in the central
gap with $\protect\varepsilon=0.01$, $k=1$, $\protect\mu=0.01$ and $\protect%
\delta=\protect\pi/2$. }
\label{f:big0m}
\end{center}
\end{figure}

To illustrate the above analysis we now introduce the Poincar\'{e} section $%
\{x=0\mod 2\pi \}$ in phase space, and consider the return map induced by
the flow of \eqref{e:ab}. We have computed the stable and unstable manifolds
of the origin using the continuation tool \texttt{AUTO} \cite{Auto} with
assistance of the \texttt{HomMap} driver \cite{Ya:98a,Ya:98b}. In
particular, to compute the stable and unstable manifolds, small segments ($%
\sim 10^{-5}$) in the stable and unstable subspaces of the origin were taken
and solutions of (\ref{e:ab}) with initial conditions on the segments were
numerically integrated. The same approach was also used for computation of
the one-dimensional unstable manifolds for a three-dimensional Poincar\'{e}
map in \cite{Ya:11}. As detected in the theory, the stable and unstable
manifolds are observed to intersect transversely and GSs could be found. The
stable and unstable manifolds on the Poincar\'{e} section $x=0\mod 2\pi $
for $\mu =0$, $\omega =\pm 0.5$, $\varepsilon =0.01$ and $k=1$ are drawn in
Fig.~\ref{f:im0}. Intersections of these manifolds give rise to homoclinic
orbits or GSs, and symmetric GSs in (\ref{e:uv}) for the same parameter
values are plotted in Fig.~\ref{f:big0}. These GSs are not visibly different
from the unperturbed ones of \eqref{e:homo0}. In Fig.~\ref{f:big0m}
asymmetric GSs in \eqref{e:uv} are also plotted for $\mu =0.01$, $\delta
=\pi /2$, $\omega =\pm 0.5$, $\varepsilon =0.01$ and $k=1$. Note that there
are different asymmetric GSs which have almost the same shapes as those in
Fig.~\ref{f:big0m} but have peaks at a position shifting by $\pi /k$. Plots
of GSs throughout this paper show the real part of the solutions as a solid
curve, whereas the imaginary part is plotted as a dashed curve.

\section{Existence of gap solitons in non-central gaps}

\label{s:av}

In order to investigate the existence of GSs of (\ref{e:uv}) with $V=-U^*$
in the non-central gaps, we shall use the higher-order averaging method \cite%
{Mu:88,Mu:91}. Let us describe a general setup for the averaging analysis.

We seek homoclinic solutions to the origin of (\ref{e:ab}) near $\omega
=\omega _{\pm m}$ in the limit of small $\varepsilon ,\mu $. Note that for $%
\varepsilon =\mu =0$ the origin is an equilibrium in (\ref{e:ab}), at which
the Jacobian matrix has a pair of purely imaginary eigenvalues $\lambda =\pm
i\sqrt{\omega ^{2}-1}$ when $\omega \approx \omega _{\pm m}$. Let $j$ and $l$
be non-negative integers which will be determined for each gap later. Define
a detuning parameter $\Omega $ by
\[
\varepsilon ^{j+1}\Omega =\omega -\omega _{\pm m}.
\]
Let
\[
\Phi (x)=%
\begin{pmatrix}
\displaystyle\cos \frac{mk}{2}x & \displaystyle-\frac{2(\omega _{\pm m}+1)}{%
mk}\sin \frac{mk}{2}x \\[2ex]
\displaystyle\frac{mk}{2(\omega _{\pm m}+1)}\sin \frac{mk}{2}x & %
\displaystyle\cos \frac{mk}{2}x%
\end{pmatrix}%
,
\]
which is the fundamental matrix of the linearization at the origin of the
unperturbed system (\ref{e:ab0}) with $\omega =\omega _{\pm m}$. Using the
transformation
\begin{equation}
\begin{pmatrix}
a \\
b%
\end{pmatrix}%
=\varepsilon ^{(l+1)/2}\,\Phi (x)%
\begin{pmatrix}
\displaystyle\frac{1}{k}\xi  \\[2ex]
\displaystyle\frac{m}{2(\omega _{\pm m}+1)}\eta
\end{pmatrix}
\label{e:trans}
\end{equation}%
in (\ref{e:ab}), one obtains
\begin{align}
&
\begin{pmatrix}
\xi ^{\prime }/k \\
(m/2)\eta ^{\prime }/(\omega _{\pm m}+1)%
\end{pmatrix}
\notag \\
& =\varepsilon \Phi ^{-1}(x)%
\begin{pmatrix}
\lbrack -\varepsilon ^{j}\Omega +\cos kx-\bar{\mu}\cos (kx+\delta )-\frac{3}{%
2}\varepsilon ^{l}(a^{2}+b^{2})]b \\[1ex]
\lbrack \varepsilon ^{j}\Omega +\cos kx+\bar{\mu}\cos (kx+\delta )+\frac{3}{2%
}\varepsilon ^{l}(a^{2}+b^{2})]a%
\end{pmatrix}%
,  \label{e:xieta}
\end{align}%
where $\bar{\mu}=\mu /\varepsilon $ and $a$ and $b$ are represented by $\xi $
and $\eta $ via (\ref{e:trans}).

Homoclinic solutions in (\ref{e:xieta}), equivalently in (\ref{e:ab}), can
be approximately obtained by the averaging method. These homoclinic
solutions correspond to GPs of (\ref{e:uv}) with $V=-U^*$ in the non-central
gaps. While in the first gaps $\mathbf{1^\pm}$ the standard (i.e.,
first-order) averaging method is sufficient, we have to use the higher-order
averaging method for the higher-order gaps. This difficulty occurs because
the potential considered here consists of the pure harmonic component $\cos
kx$. But it is important to note that higher-order averaging is also needed
for more general potentials with higher Fourier components, if these higher
components are small compared to the first one. This is the reason for us to
present the higher-order method here.

The necessary lengthy computations can be easily performed with computer
algebra systems. Specifically, we have used the program \texttt{haverage.m}
for the computer software \texttt{Mathematica} \cite{Wo:03}, developed in
\cite{Ya:99,YaIc:99} (see also \cite{Ya:98c}). We exemplarily present the
results for the first three gaps $\mathbf{1^\pm}$, $\mathbf{2^\pm}$ and $%
\mathbf{3^\pm}$. Similar second-order averaging analyses can also calculate
the boundaries of the central gap (see \ref{a:bound}).

\subsection{Averaging in the gaps $\mathbf{1^\pm}$}

For the computations in the first gap at $\omega_{\pm 1}=\pm\sqrt{1+k^2/4}$,
we set $m=1$ and $j=l=0$. The standard averaging procedure for (\ref{e:xieta}%
) yields the first-order averaged system
\begin{equation}
\begin{split}
& \xi^{\prime}=\varepsilon\left[ -\frac{\bar{\mu}\sin\delta}{k}\xi +\frac{%
\omega_{\pm 1}(1-2\Omega)+\bar{\mu}\cos\delta}{k}\eta -\gamma_{\pm
1}(\xi^2+\eta^2)\eta\right], \\
& \eta^{\prime}=\varepsilon\left[ \frac{\omega_{\pm 1}(1+2\Omega)+\bar{\mu}%
\cos\delta}{k}\xi +\frac{\bar{\mu}\sin\delta}{k}\eta +\gamma_{\pm
1}(\xi^2+\eta^2)\xi\right],
\end{split}
\label{e:av1}
\end{equation}
where
\[
\gamma_{\pm 1}=\frac{3(2\omega_{\pm 1}^2+1)}{2k^3(\omega_{\pm 1}+1)^3}>0.
\]
Eqs.~\eqref{e:av1} are easily obtained by averaging the right hand side of %
\eqref{e:xieta} with respect to $x$ over $[0,2\pi/k]$ and such computations
can be done by the \texttt{Mathematica} program \texttt{haverage.m}. In the
averaged system (\ref{e:av1}) with $\bar{\mu}\cos\delta+\omega_{\pm 1}\neq 0$%
, we perform a rotational transformation
\[
\begin{pmatrix}
\tilde{\xi} \\
\tilde{\eta}%
\end{pmatrix}
=
\begin{pmatrix}
\cos\theta & \sin\theta \\
-\sin\theta & \cos\theta%
\end{pmatrix}
\begin{pmatrix}
\xi \\
\eta%
\end{pmatrix}
\]
to obtain a system of a more convenient form
\begin{equation}
\begin{split}
& \tilde{\xi}^{\prime}=\varepsilon\left[ \frac{2\omega_{\pm 1}}{k}%
(\Omega_{1\pm}-\Omega)\tilde{\eta} -\gamma_{\pm 1}(\tilde{\xi}^2+\tilde{\eta}%
^2)\tilde{\eta}\right], \\
& \tilde{\eta}^{\prime}=\varepsilon\left[ \frac{2\omega_{\pm 1}}{k}%
(\Omega_{1\pm}+\Omega)\tilde{\xi} +\gamma_{\pm 1}(\tilde{\xi}^2+\tilde{\eta}%
^2)\tilde{\xi}\right],
\end{split}
\label{e:av1a}
\end{equation}
where
\[
\Omega_{1\pm} =\frac{1}{2} \sqrt{\left(\frac{\bar{\mu}}{\omega_{\pm 1}}%
\right)^2 +2\left(\frac{\bar{\mu}}{\omega_{\pm 1}}\right)\cos\delta+1}
\]
and
\[
\theta=
\begin{cases}
\theta_1 & \mbox{for $\omega_{\pm 1}(\bar{\mu}\cos\delta+\omega_{\pm 1})>0$};
\\
\displaystyle \theta_1+\frac{\pi}{2} &
\mbox{for $\omega_{\pm
1}(\bar{\mu}\cos\delta+\omega_{\pm 1})<0$}%
\end{cases}%
\]
with
\[
\theta_1=\frac{1}{2}\arctan \left(\frac{\bar{\mu}\sin\delta} {\bar{\mu}%
\cos\delta+\omega_{\pm 1}}\right)\in[-\pi,\pi].
\]
Note that Eq.~(\ref{e:av1a}) is still valid for $\bar{\mu}%
\cos\delta+\omega_{\pm 1}=0$ if $\theta=\pi/4$ is taken when $\omega_{\pm 1}%
\bar{\mu}\sin\delta>0$ and if $\theta=-\pi/4$ is taken when $\omega_{\pm 1}%
\bar{\mu}\sin\delta<0$. The system (\ref{e:av1a}) is Hamiltonian with a
Hamilton function
\begin{align*}
H_1(\tilde{\xi},\tilde{\eta}) =& \varepsilon\left\{ \frac{\omega_{\pm 1}}{k}%
\left[(\Omega_{1\pm}+\Omega)\tilde{\xi}^2 -(\Omega_{1\pm}-\Omega)\tilde{\eta}%
^2\right] +\frac{1}{4}\gamma_{\pm 1}(\tilde{\xi}^2+\tilde{\eta}^2)^2\right\},
\end{align*}
for which the origin $(\tilde{\xi},\tilde{\eta})=(0,0)$ is a saddle if $%
\Omega\in(-\Omega_{1\pm},\Omega_{1\pm})$, i.e.,
\begin{equation}
\omega\in\left(\omega_{\pm 1}-\varepsilon\Omega_{1\pm}, \omega_{\pm
1}+\varepsilon\Omega_{1\pm}\right).  \label{e:region1}
\end{equation}
We easily see that if
\begin{equation}
\mu=\varepsilon\sqrt{1+\frac{1}{4}k^2},\quad \delta=0\quad%
\mbox{(resp.
$\delta=\pi$)}  \label{e:closing}
\end{equation}
then $\Omega_{1+}=0$ (resp. $\Omega_{1-}=0$) {and a closing of the gap
region $\mathbf{1^+}$ (resp. $\mathbf{1^-}$) occurs. See also Fig.~\ref%
{f:linear}(d).} If $\delta\neq 0,\pi$, then $\Omega_{1\pm}\neq 0$, and thus
such a closing does not occur.

Let us assume that $\Omega_{1\pm}\neq 0$, i.e., condition~(\ref{e:closing})
does not hold. A straightforward analysis of the level sets of $H_1$ shows
the existence of two homoclinic orbits in (\ref{e:av1a}) and hence in (\ref%
{e:av1}). Moreover, we can obtain analytical expressions of the homoclinic
orbits as
\begin{align}
& (\tilde{\xi}_\pm(x),\tilde{\eta}_\pm(x))  \notag \\
&=\left( \pm 2\sqrt{\frac{2\omega_{\pm 1}\Omega_{1\pm}}{3k\gamma_{\pm 1}}}%
\sin\phi \sin\frac{1}{2}\phi\, \frac{\sinh[(2\omega_{\pm
1}\Omega_{1\pm}\sin\phi/k)\varepsilon x]} {\cosh[(4\omega_{\pm
1}\Omega_{1\pm}\sin\phi/k)\varepsilon x]+\cos\phi}, \right.  \notag \\
& \qquad \left. \mp 2\sqrt{\frac{2\omega_{\pm 1}\Omega_{1\pm}}{3k\gamma_{\pm
1}}}\sin\phi \cos\frac{1}{2}\phi\, \frac{\cosh[(2\omega_{\pm 1}\Omega_{1\pm}
\sin\phi/k)\varepsilon x]} {\cosh[(4\omega_{\pm 1}\Omega_{1\pm}
\sin\phi/k)\varepsilon x]+\cos\phi} \right),  \label{e:homo}
\end{align}
where $\Omega/\Omega_{1\pm}=\cos\phi$.

It is easy to see that, since the coefficient in front of $\tilde{\eta}^2$
in $H_1$ is negative, the homoclinic orbits intersect the $\tilde{\eta}$%
-axis and are symmetric about it, forming a figure-of-eight configuration.
In general, it is very difficult to make precise statements about the
symmetry of homoclinic orbits in (\ref{e:av1}). For some special cases,
however, we have the following results:

\begin{enumerate}
\renewcommand{\labelenumi}{(\roman{enumi})}
\item When $\bar{\mu}>0$ and $\delta=0$, the homoclinic orbits in the
averaged system (\ref{e:av1}) are symmetric about the $\eta$-axis if $\bar{%
\mu}<-\omega_{\pm 1}$ and about the $\xi$-axis if $\bar{\mu}>-\omega_{\pm 1}
$ in the negative gap, while they are always symmetric about the $\eta$-axis
in the positive gap.

\item When $\mu>0$ and $\delta=\pi$, the homoclinic orbits in (\ref{e:av1})
are symmetric about the $\eta$-axis if $\bar{\mu}<\omega_{\pm 1}$ and about
the $\xi$-axis if $\bar{\mu}>\omega_{\pm 1}$ in the positive gap, while they
are always symmetric about the $\eta$-axis in the negative gap.

\item When $\mu=0$, the homoclinic orbits in (\ref{e:av1}) are symmetric
about the $\eta$-axis in both the positive and negative gaps.
\end{enumerate}

From the existence of homoclinic orbits in the averaged system \eqref{e:av1}%
, we claim that there exist homoclinic orbits in \eqref{e:u} and hence
stationary GSs in \eqref{e:uv}. For this consider (\ref{e:xieta}) and (\ref%
{e:av1}) as dynamical systems defined in the phase space $\mathbb{R}^2\times%
\mathbb{S}^1$. Recall that we set $m=1$ and $j=l=0$ in (\ref{e:xieta}). We
introduce a cross section $\Sigma$ at $x=0\mod 2\pi/k$ and define the Poincar%
\'{e} maps $\psi,\bar{\psi}:\Sigma\rightarrow\Sigma$ for (\ref{e:xieta}) and
(\ref{e:av1}), respectively. For both $\psi$ and $\bar{\psi}$ the origin $%
(\xi,\eta)=(0,0)$ is a fixed point and has stable and unstable manifolds.
Moreover, the stable and unstable manifolds, $\bar{W}^s(0)$ and $\bar{W}%
^u(0) $, coincide along the homoclinic orbits in (\ref{e:av1}) for $\bar{\psi%
}$. By the averaging theorem (see Theorem~4.1.1 of \cite{GuHo:83}), the
stable and unstable manifolds, $W^s(0)$ and $W^u(0)$, for $\psi$ are $%
O(\varepsilon)$-close to $\bar{W}^s(0)$ and $\bar{W}^u(0)$. Since Eq.~(\ref%
{e:xieta}) as well as Eq.~(\ref{e:av1}) are Hamiltonian, the Poincar\'{e}
maps $\psi,\bar{\psi}$ are area-preserving.

\begin{figure}[tbp]
\begin{center}
\includegraphics[scale=0.9]{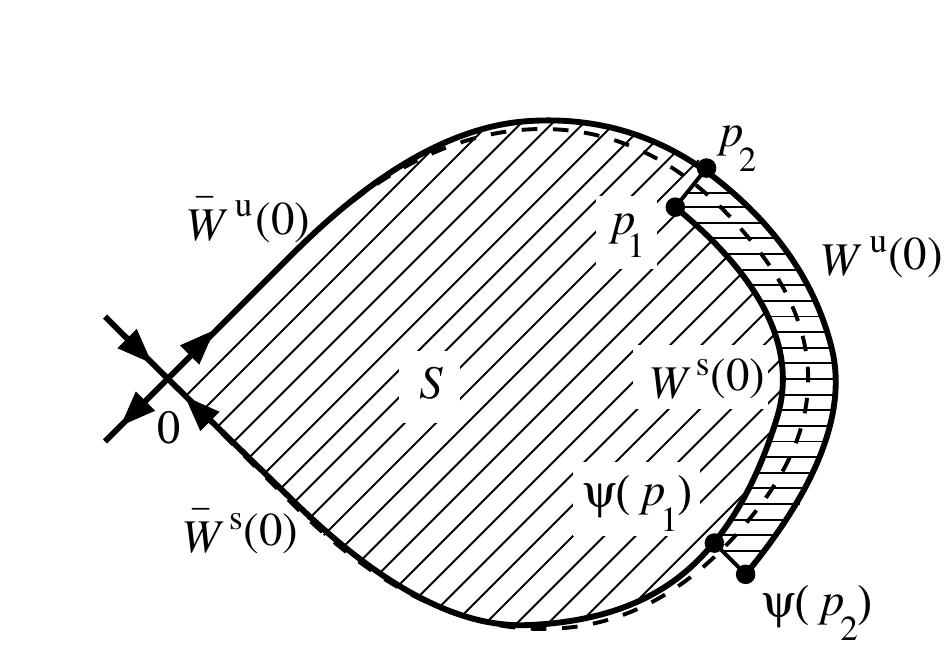}
\caption{Stable and unstable manifolds $W^{s,u}(0)$ (solid lines) which are
close to $\bar{W}^{s,u}(0)$ (broken lines) but do not intersect. The regions
$S$ and $S^{\prime}\backslash S$ with $S^{\prime}=\protect\psi(S)$ are
hatched distinctly.}
\label{f:asu}
\end{center}
\end{figure}

Now assume that the stable and unstable manifolds $W^{s}(0)$ and $W^{u}(0)$
do not intersect, and go inward and outward, respectively, as shown in Fig.~%
\ref{f:asu}. We want to show that this yields a contradiction and define a
region $S$, which is encircled by two parts of $W^{s}(0)$ and $W^{u}(0)$ and
a line connecting two points $p_{1}$ and $p_{2}$ on $W^{s}(0)$ and $W^{u}(0)$%
. The region $S$ is mapped by the Poincar\'{e} map $\psi $ to another region
$S^{\prime }=\psi (S)$ which is encircled by two different parts of $W^{s}(0)
$ and $W^{u}(0)$ and a curve connecting two points $\psi (p_{1})$ and $\psi
(p_{2})$ on $W^{s}(0)$ and $W^{u}(0)$. It is obvious that $S^{\prime }$ is
larger than $S$ and contradicts the fact that $\psi $ is area-preserving.
Similarly, we cannot assume that $W^{s}(0)$ and $W^{u}(0)$ go outward and
inward, respectively. Therefore, we can prove that $W^{s}(0)$ and $W^{u}(0)$
intersect and there exists a homoclinic orbit in the full system (\ref%
{e:xieta}) and hence in (\ref{e:u}). On the other hand, Eq.~\eqref{e:xieta}
can be averaged up to $O(\varepsilon ^{n})$ for any integer $n\geq 1$ but
has very rapid oscillations compared with the unperturbed homoclinic orbits %
\eqref{e:homo} which vary very slowly with $O(1/\varepsilon )$. Applying a
result by Neishtadt \cite{Ne:84} (see also \cite{Si:94}), we see that the
splitting distance between $W^{s}(0)$ and $W^{u}(0)$ is exponentially small
with respect to $\varepsilon $ at most. Thus, there exist stationary GS
solutions to the system of equations (\ref{e:uv}) in the gap regions %
\eqref{e:region1}.

Moreover, assume that $\mu\sin\delta=0$, i.e., $\mu=0$ or $\delta=0,\pi$, so
that Eqs.~(\ref{e:ab}) is reversible with respect to $R_j$ and $R_j^{\prime}$%
, $j=1,2$. Then it follows from the above arguments that the homoclinic
orbit in the averaged system (\ref{e:av1}) is symmetric about the $\xi$-axis
(resp. $\eta$-axis). By the persistence of symmetric orbits in reversible
systems, there exist symmetric homoclinic orbits about the $\xi$-axis (resp.
$\eta$-axis) in the full system (\ref{e:xieta}) and with respect to $R_1$
and $R_1^{\prime}$ (resp. $R_2$ and $R_2^{\prime}$) in (\ref{e:ab}) and
hence in (\ref{e:u}). In particular, the homoclinic orbits in (\ref{e:u})
have an even (resp. odd) real part $a$ and odd (resp. even) imaginary part $%
b $.

We can use similar arguments to the central gap when condition~(\ref{e:con})
does not hold, and prove that there must be homoclinic orbits. However, the
splitting distance between the stable and unstable manifolds is not
exponentially small with respect to $\varepsilon$ but $O(\varepsilon^2)$ at
most since both of the perturbations in \eqref{e:ab} and unperturbed
homoclinic orbits \eqref{e:homo0} vary with $O(1)$.

\begin{figure}[t]
\begin{center}
\includegraphics[scale=0.9]{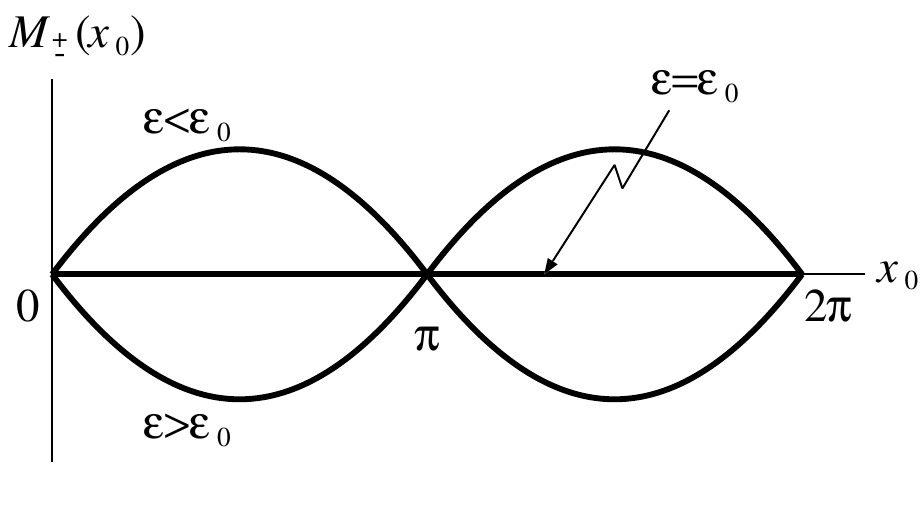}
\caption{Graphs of the Melnikov functions $M_\pm(x_0)$ for $\delta=\pi$.}
\label{f:Mel}
\end{center}
\end{figure}

\begin{figure}[t]
\begin{center}
\includegraphics[scale=0.6]{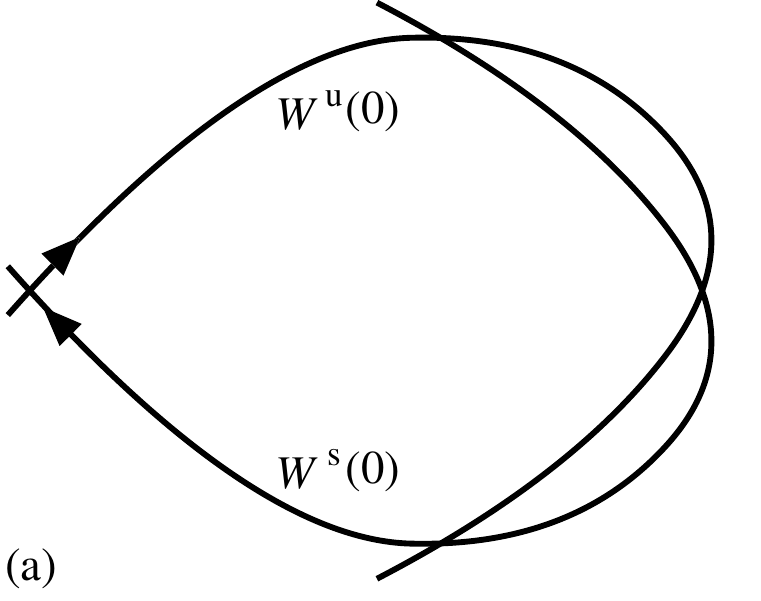}\qquad %
\includegraphics[scale=0.6]{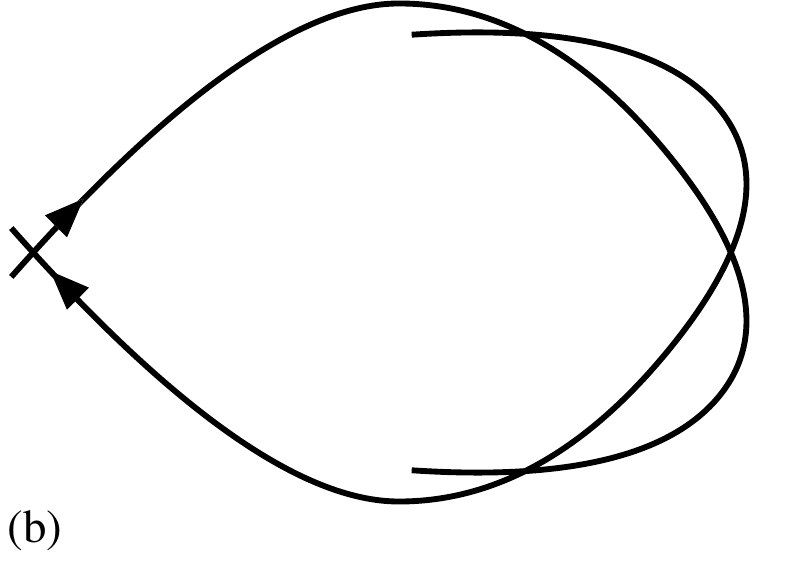}\\[2ex]
\includegraphics[scale=0.6]{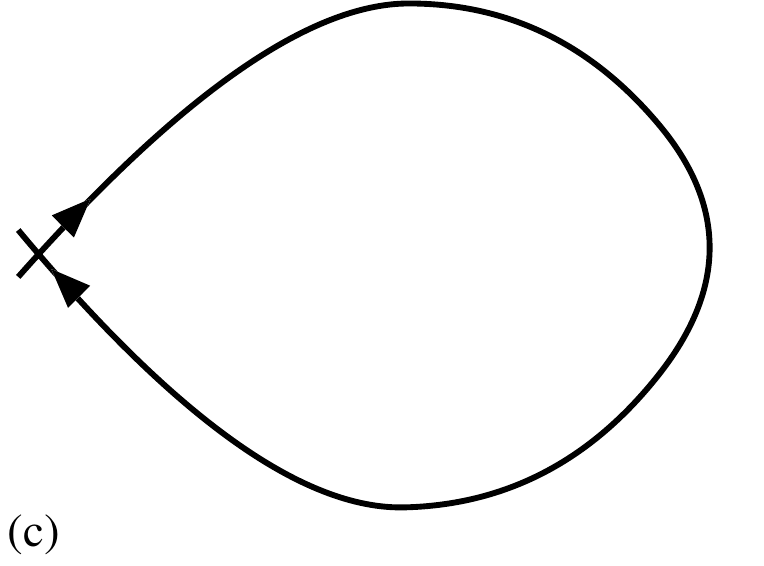}
\caption{Stable and unstable manifolds $W^{s,u}(0)$ for the associated
Poincar\'{e} map: (a) $\protect\varepsilon<\protect\varepsilon_0$; (b) $%
\protect\varepsilon>\protect\varepsilon_0$; (c) $\protect\varepsilon=\tilde{%
\protect\varepsilon}_0$.}
\label{f:su}
\end{center}
\end{figure}

In addition, we can detect a non-transverse homoclinic orbit in (\ref{e:ab})
when $\delta=\pi$ and
\begin{equation}
\varepsilon\approx\varepsilon_0 =\frac{2\mu}{k}\tanh\left(\frac{k\theta}{%
2\sin\theta}\right),  \label{eqn:hbc}
\end{equation}
as follows. Fix $\delta=\pi$. Then the Melnikov functions $M_\pm(x_0)$ have
graphs as shown in Fig.~\ref{f:Mel}, depending on $\varepsilon<\varepsilon_0$%
, $\varepsilon=\varepsilon_0$ or $\varepsilon>\varepsilon_0$. Since $%
M_\pm(x_0) $ represent signed measures of their distance (see, e.g.,
section~4.5 of \cite{Wi:90}), noting that $b_\pm(x_0)=0$ and Eqs.~(\ref{e:ab}%
) is reversible, we can draw the behavior of the stable and unstable
manifolds $W^{s,u}(0)$ for the Poincar\'{e} map of (\ref{e:ab}) as shown in
Figs.~\ref{f:su}(a) and (b). Using the fact that the Poincar\'{e} map is
area-preserving and applying a discussion similar to the above one for (\ref%
{e:xieta}), we see that $W^s(0)$ and $W^u(0)$ coincide for some $\varepsilon$
near $\varepsilon_0$, say $\tilde{\varepsilon}_0$, as shown in Fig.~\ref%
{f:su}(c). Thus, the point $(\delta,\varepsilon)=(\pi,\tilde{\varepsilon}_0)$
is very degenerate in the parameter space. This also suggests that new
interesting behavior may occur at the point in the original PDEs (\ref{e:uv}%
).

\begin{figure}[t]
\begin{center}
\includegraphics[scale=0.65]{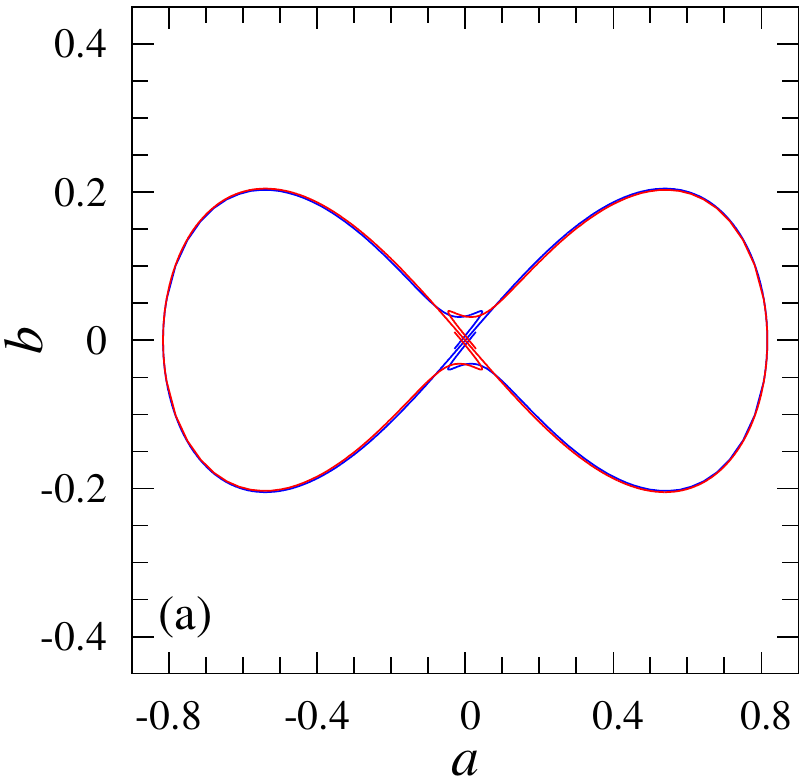}\qquad %
\includegraphics[scale=0.65]{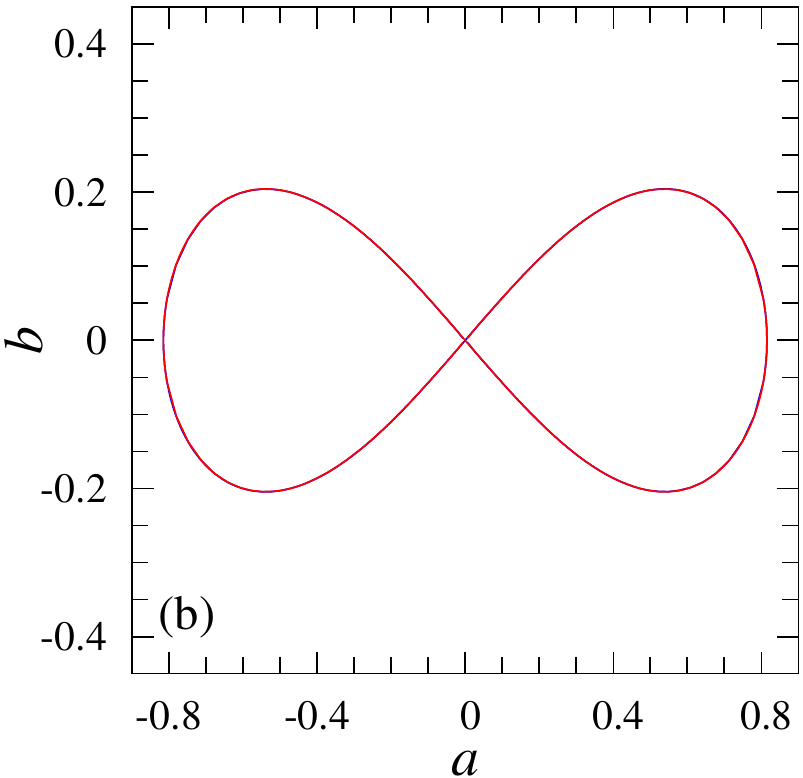}\\[2ex]
\includegraphics[scale=0.65]{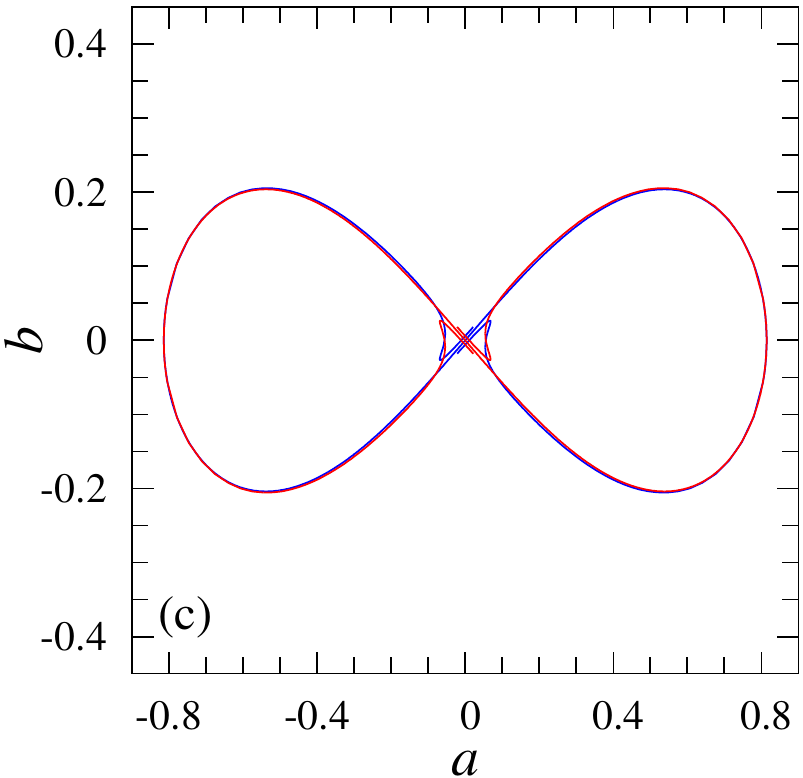}
\caption{Stable and unstable manifolds of the origin on the Poincar\'{e}
section $x=0\mod 2\protect\pi$ when $\protect\mu=0.01$, $\protect\omega=0.5$%
, $\protect\delta=\protect\pi$ and $k=1$: (a) $\protect\varepsilon=0.0088063$%
; (b) $\protect\varepsilon=0.0108063$; (c) $\protect\varepsilon=0.0128063$.
The red and blue curves represent the stable and unstable manifolds,
respectively.}
\label{f:im0'}
\end{center}
\end{figure}

\begin{figure}[t]
\begin{center}
\includegraphics[scale=0.65]{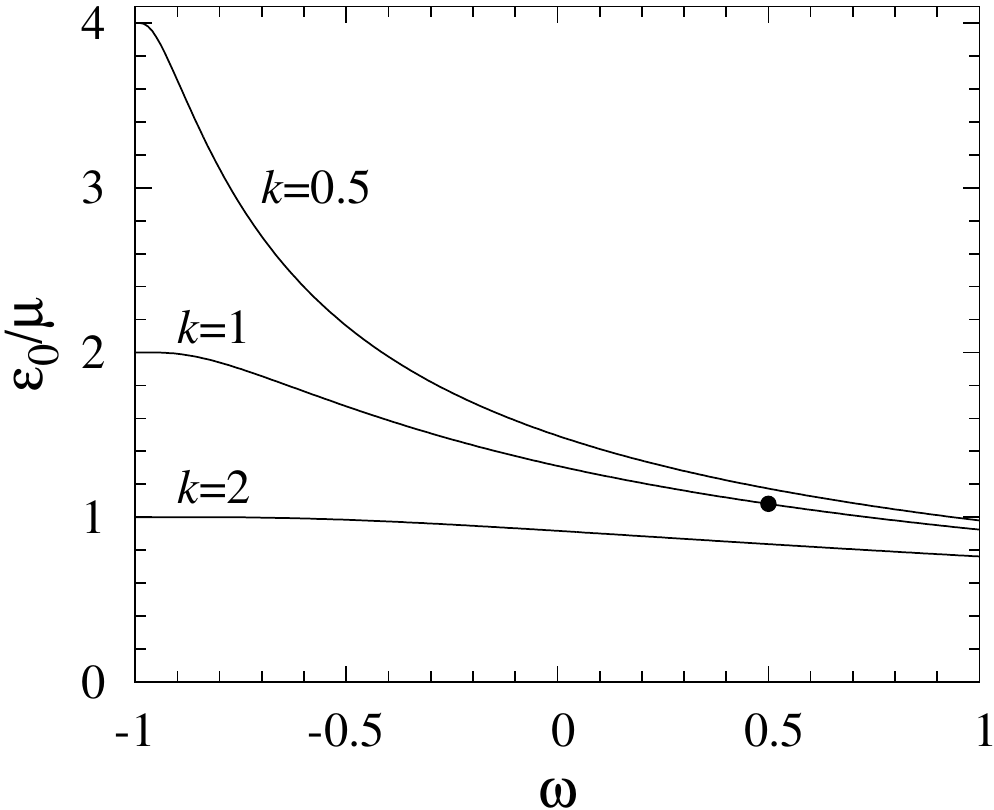}
\caption{Homoclinic bifurcation curves for Eqs.~\eqref{e:ab} with $k=0.5,1,2$.
The parameter values $(\omega,\epsilon)$ of Fig.~\ref{f:im0'}(b) with $k=1$
 is plotted as $\bullet$.}
\label{f:hbc}
\end{center}
\end{figure}

In Fig.~\ref{f:im0'} we show the results of numerical computations of the
stable and unstable manifolds by \texttt{AUTO97} with \texttt{HomMap} near
the homoclinic bifurcation point $\varepsilon=\varepsilon_0\approx 0.010806$
for $\mu=0.01$, $\omega=0.5$ and $k=1$. We see that as predicted by the
theory, the invariant manifolds almost coincide at $\varepsilon\approx%
\varepsilon_0$ while they split when $\varepsilon$ is different from $%
\varepsilon_0$. Thus, the theoretical prediction for the homoclinic
bifurcations is very precise. Figure~\ref{f:hbc} also shows the approximate
homoclinic bifurcation curves given by Eq. \eqref{eqn:hbc}, on which the
stable and unstable manifolds coincide, in the $(\omega,\varepsilon/\mu)$%
-space for $k=0.5,1,2$.

\subsection{Averaging in the gaps $\mathbf{2^\pm}$}

We set $m=2$ and $j=l=1$, and perform a second-order averaging method to
prove the existence of gap solitons of (\ref{e:uvans}) with $V=-U^*$ in the
second gaps appearing at $\omega_{\pm 2}=\pm \sqrt{1+k^2}$. To avoid the
complexity of expressions, we give our results only for $\mu=0$ in the gaps
of higher orders. The second-order averaged system has been obtained using
the \texttt{Mathematica} program \texttt{haverage.m}, and it reads
\begin{equation}
\begin{split}
\xi^{\prime}=& \displaystyle{-\frac{\varepsilon^{2}}{k^{2}}} \left\{
\alpha_{\pm 2} \left[(6(\omega_{\pm 2}+1)\Omega-5)k^{2}+ (\omega_{\pm
2}+1)^{2}\right] \right. \\
& \qquad +\beta_{\pm 2}\left[k^{2}\xi^{2}+ (\omega_{\pm 2}+1)^{2} \eta^{2}%
\right]\big\}\; \eta, \\
\eta^{\prime}= & \displaystyle{\frac{\varepsilon^{2}}{(\omega_{\pm 2}+1)^{2}}%
} \left\{\alpha_{\pm 2} \left[(6(\omega_{\pm 2}+1)\Omega+1)k^{2}-
5(\omega_{\pm 2}+1)^{2}\right] \right. \\
& \qquad +\beta_{\pm 2}\left[k^{2}\xi^{2} +(\omega_{\pm 2}+1)^{2}\eta^{2}%
\right]\big\} \; \xi,
\end{split}
\label{e:av2}
\end{equation}
with
\begin{align*}
\alpha_{\pm 2} =& \frac{k^{2}+(\omega_{\pm 2}+1)^{2}}{12k^{2}(\omega_{\pm
2}+1)}, \\
\beta_{\pm 2} =& \frac{3[3k^{4}+2(\omega_{\pm 2}+1)^{2}k^{2}+3(\omega_{\pm
2}+1)^{4}]} {16k^{2}(\omega_{\pm 2}+1)^{2}}.
\end{align*}
Note that $\alpha_{+2}=-\alpha_{-2}>0$ and $\beta_{\pm 2}>0$. Eq.~(\ref%
{e:av2}) is a Hamiltonian system with the Hamilton function
\begin{align*}
H_2(\xi,\eta) =& \varepsilon^{2}\biggl( \frac{\alpha_{\pm 1/1}
\{[6(\omega_{\pm 1/1}+1)\Omega+1]k^{2}-5(\omega_{\pm 1/1}+1)^{2}\}} {%
2(\omega_{\pm 1/1}+1)^{2}}\xi^{2}  \notag \\
& \quad +\frac{\alpha_{\pm 1/1} \{[6(\omega_{\pm
1/1}+1)\Omega-5]k^{2}+(\omega_{\pm 1/1}+1)^{2}\}} {2k^{2}}\eta^{2}  \notag \\
& \quad +\frac{\beta_{\pm 1/1}}{4k^{2}(\omega_{\pm 2/1}+1)^{2}}
[k^{2}\xi^{2}+(\omega_{\pm 1/1}+1)^{2}\eta^{2}]^{2}\biggr).
\end{align*}

Let
\begin{align*}
& \Delta\Omega_{2}^{1} =\frac{2k^{2}-\sqrt{k^{2}+1}-1}{3k^{2}(\sqrt{k^{2}+1}%
+1)} =\frac{2k^{2}-5\sqrt{k^{2}+1}+5}{3k^{2}(\sqrt{k^{2}+1}-1)}, \\
& \Delta\Omega_{2}^{2} =\frac{2k^{2}+5\sqrt{k^{2}+1}+5}{3k^{2}(\sqrt{k^{2}+1}%
+1)} =\frac{2k^{2}+\sqrt{k^{2}+1}-1}{3k^{2}(\sqrt{k^{2}+1}-1)}.
\end{align*}
The origin is a saddle of (\ref{e:av2}) and has a pair of homoclinic orbits
if $\Omega\in(\Delta\Omega_{2}^{1},\Delta\Omega_{2}^{2})$ and $%
\Omega\in(-\Delta\Omega_{2}^{2},-\Delta\Omega_{2}^{1})$ when the signs `$+$'
and `$-$' are taken in the subscript of $\omega$ in (\ref{e:av2}),
respectively, i.e.,
\[
\omega\in \left(\omega_{+2}+\varepsilon^{2}\Delta\Omega_{2}^{1},
\omega_{+2}+\varepsilon^{2}\Delta\Omega_{2}^{2}\right)
\]
or
\[
\omega\in \left(\omega_{-2}-\varepsilon^{2}\Delta\Omega_{2}^{2},
\omega_{-2}-\varepsilon^{2}\Delta\Omega_{2}^{1}\right).
\]

Again, we find the averaged system to be Hamiltonian with a saddle
equilibrium at the origin, and a study of the corresponding level set of the
Hamiltonian shows the existence of a figure-of-eight of homoclinic orbits.
These orbits intersect the $\eta$-axis. By the same argument as above, this
implies the existence of gap solitons for (\ref{e:u}) in the gaps $\mathbf{%
2^\pm}$ having an even $a$-component and an odd $b$-component.

\subsection{Averaging in the gaps $\mathbf{3^\pm}$}

We set $m=3$, $j=1$ and $l=2$, and perform the third-order averaging
procedure for (\ref{e:xieta}) using the \texttt{Mathematica} program \texttt{%
haverage.m} to obtain the averaged system
\begin{equation}
\begin{split}
\xi^{\prime}=& -\frac{\varepsilon^{2}}{9k^{2}}\rho_{\pm 3}
\left\{[32(\omega_{\pm 3}+1)\Omega-9]k^{2} -4(\omega_{\pm
3}+1)^{2}\right\}\eta \\
& +\frac{\varepsilon^{3}}{9k^{2}} \left\{\alpha_{\pm 3}-\beta_{\pm 3}\left[%
9k^{2}\xi^{2} +4(\omega_{\pm 3}+1)^{2}\eta^{2}\right]\right\}\eta, \\
\eta^{\prime}=& \frac{\varepsilon^{2}}{4(\omega_{\pm 3}+1)^{2}}\rho_{\pm 3}
\left\{[32(\omega_{\pm 3}+1)\Omega-9]k^{2} -4(\omega_{\pm
3}+1)^{2}\right\}\xi \\
& +\frac{\varepsilon^{3}}{4(\omega_{\pm 3}+1)^{2}} \left\{\alpha_{\pm
3}+\beta_{\pm 3}\left[9k^{2}\xi^{2} +4(\omega_{\pm 3}+1)^{2}\eta^{2}\right]%
\right\}\xi,
\end{split}
\label{e:av3o}
\end{equation}
where
\begin{align*}
\rho_{\pm 3} =& \frac{9k^{2}+4(\omega_{\pm 3}+1)^{2}}{64k^{2}(\omega_{\pm
3}+1)}, \\
\alpha_{\pm 3} =& \frac{3[9k^{2}+4(\omega_{\pm 3}+1)^{2}]
[81k^{4}-56(\omega_{\pm3}+1)^{2}k^{2}+16(\omega_{\pm3}+1)^{4}]} {%
512k^{4}(\omega_{\pm 3}+1)^{2}}, \\
\beta_{\pm 3} =& \frac{81k^{4}+24(\omega_{\pm 3}+1)^{2}k^{2}+16(\omega_{\pm
3}+1)^{4}} {64k^{2}(\omega_{\pm 3}+1)^{2}}.
\end{align*}
Note that $\rho_{+3}=-\rho_{-3}>0$ and $\alpha_{\pm 3},\beta_{\pm 3}>0$
since
\begin{align*}
& 81k^{4}-56(\omega_{\pm3}+1)^{2}k^{2}+16(\omega_{\pm3}+1)^{4} \\
& =[9k^{2}-4(\omega_{\pm3}+1)^{2}]^{2}+18(\omega_{\pm3}+1)^{2}>0.
\end{align*}

Let
\[
\Omega=\Omega_{\pm 3}+\varepsilon\nu,
\]
where
\[
\Omega_{\pm 3} =\frac{9k^{2}+4(\omega_{\pm 3}+1)^{2}}{32(\omega_{\pm
3}+1)k^{2}}.
\]
Note that $\Omega_{+3}>0$ and $\Omega_{-3}<0$. Eq.~(\ref{e:av3o}) becomes
\begin{equation}
\begin{split}
\xi^{\prime}=& -\frac{\varepsilon^{3}}{9k^{2}} \{(\gamma_{\pm
3}\nu-\alpha_{\pm 3}) +\beta_{\pm 3}[9k^{2}\xi^{2}+4(\omega_{\pm
3}+1)^{2}\eta^{2}]\}\eta, \\
\eta^{\prime}=& \frac{\varepsilon^{3}}{4(\omega_{\pm 3}+1)^{2}}\{
(\gamma_{\pm 3}\nu+\alpha_{\pm 3}) +\beta_{\pm
3}[9k^{2}\xi^{2}+4(\omega_{\pm 3}+1)^{2}\eta^{2}]\}\xi,
\end{split}
\label{e:av3}
\end{equation}
where
\[
\gamma_{\pm 3}=32\rho_{\pm 3}(\omega_{\pm 3}+1)k^{2} =\frac{1}{2}%
[9k^{2}+4(\omega_{\pm 3}+1)^{2}]>0.
\]
Eq.~(\ref{e:av3}) is a Hamiltonian system with the Hamilton function
\begin{align}
H_3(\xi,\eta) =& \varepsilon^{3}\biggl( \frac{\gamma_{\pm 3}\nu+\alpha_{\pm
3}}{8(\omega_{\pm 3}+1)^{2}}\xi^{2} +\frac{\gamma_{\pm 3}\nu-\alpha_{\pm 3}}{%
18k^{2}}\eta^{2}  \notag \\
& +\frac{\beta_{\pm 3}}{144k^{2}(\omega_{\pm 2/1}+1)^{2}} [9k^{2}\xi^{2}+4(%
\omega_{\pm 2/1}+1)^{2}\eta^{2}]^{2}\biggr).  \label{e:H3}
\end{align}

We easily see that
\[
\frac{\alpha_{+3}}{\gamma_{+3}} =\frac{\alpha_{-3}}{\gamma_{-3}} =\frac{%
3(k^{4}+4)}{16k^{4}}
\]
and set
\[
\Delta\Omega_{3}=\frac{3(k^{4}+4)}{16k^{4}}.
\]
The origin is a saddle and has a pair of homoclinic orbits if $%
\nu\in(-\Delta\Omega_{3},\Delta\Omega_{3})$, i.e.,
\[
\omega\in\left( \omega_{\pm 3}\pm\varepsilon^{2}\Omega_{3}
-\varepsilon^{3}\Delta\Omega_{3}, \omega_{\pm 3}\pm\varepsilon^{2}\Omega_{3}
+\varepsilon^{3}\Delta\Omega_{3} \right).  \label{e:region3}
\]
Especially, in the regions (\ref{e:region3}) the coefficients of $\xi^{2}$
and $\eta^{2}$ in the Hamiltonian (\ref{e:H3}) are positive and negative,
respectively, so that the pair of homoclinic orbits draws a vertical
figure-eight' in the $(\xi,\eta)$-phase plane.

Similar computations reveal the existence of gap solitons in the higher
order gaps. These solutions are odd in $a$ (and even in $b$) or vice versa,
depending on whether the gap number is odd or even. A general proof of this
is beyond the scope of this paper. In the following we illustrate and extend
this perturbative results by numerical computations.

\subsection{Numerical studies}

\begin{figure}[t]
\begin{center}
\includegraphics[scale=0.65]{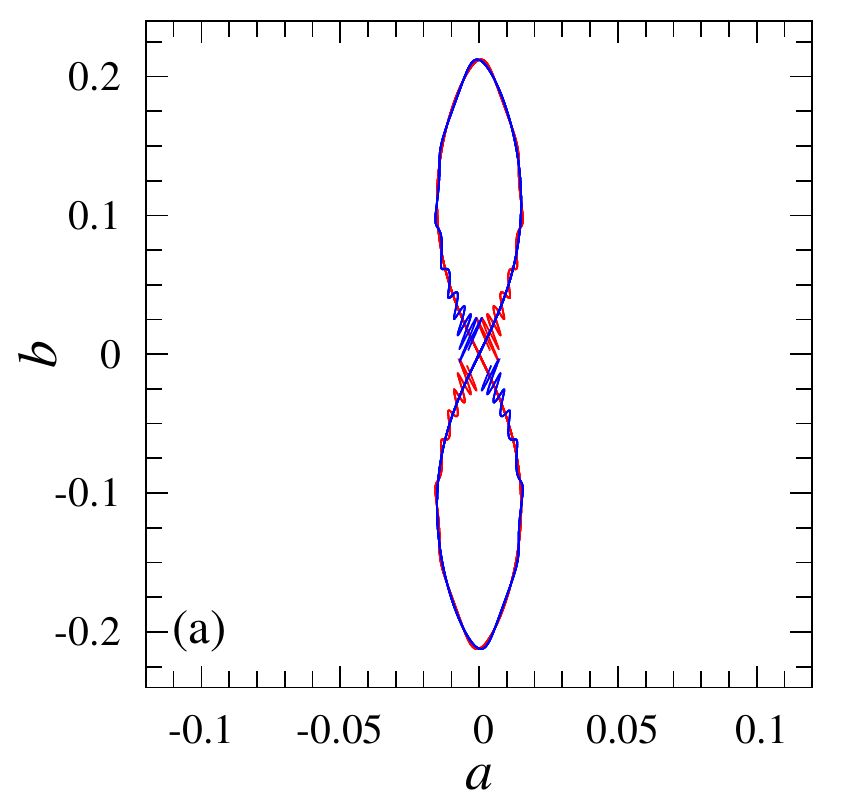}\qquad %
\includegraphics[scale=0.65]{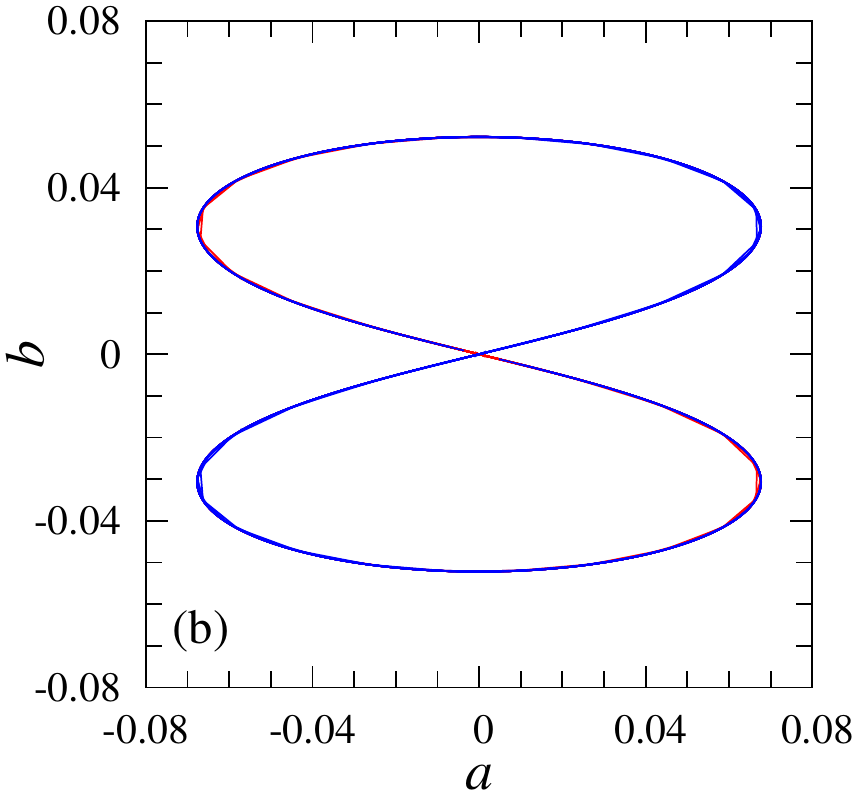}\\[3ex]
\includegraphics[scale=0.65]{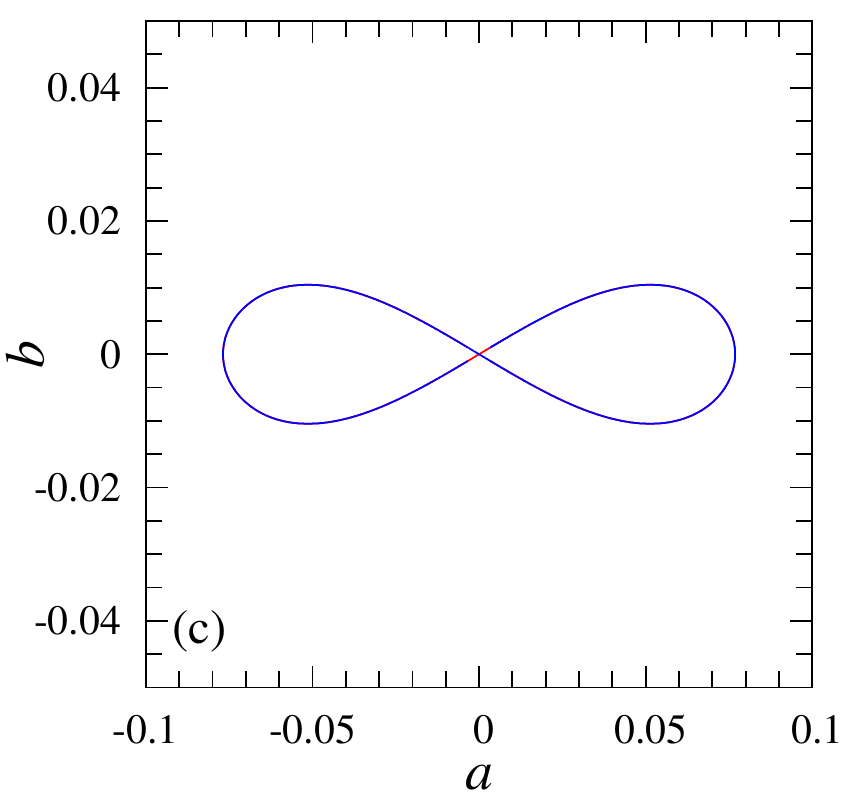}\qquad %
\includegraphics[scale=0.65]{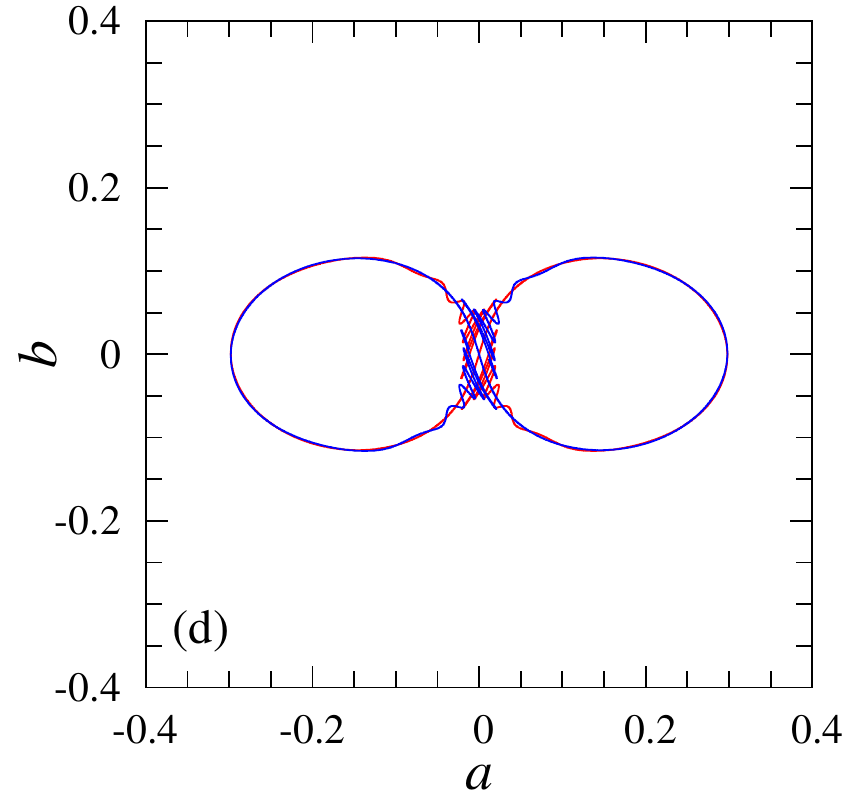}\\[3ex]
\includegraphics[scale=0.65]{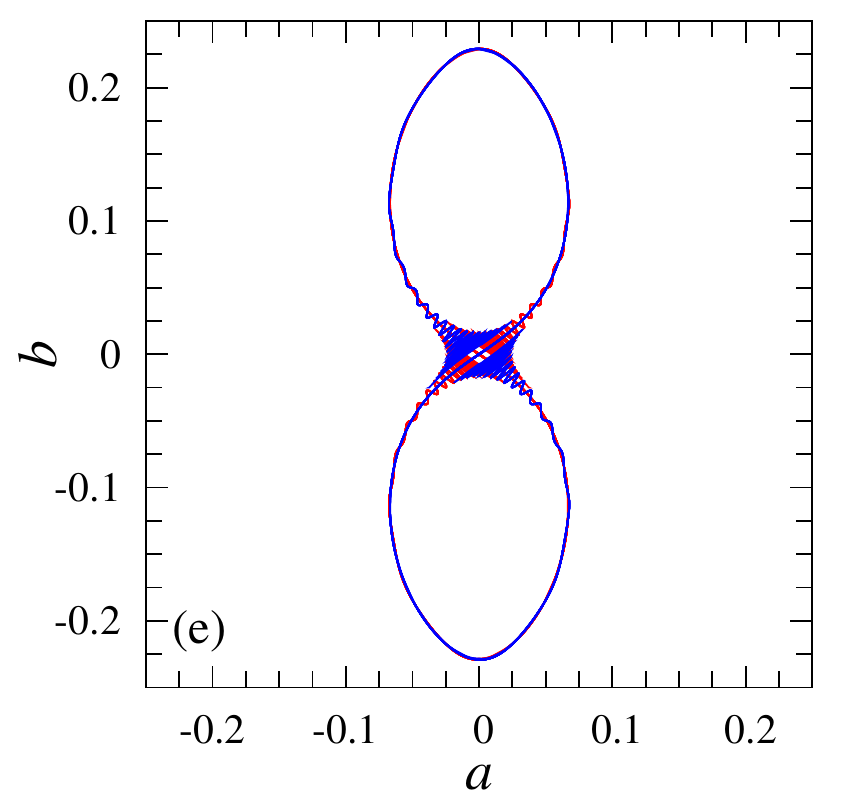}\qquad %
\includegraphics[scale=0.65]{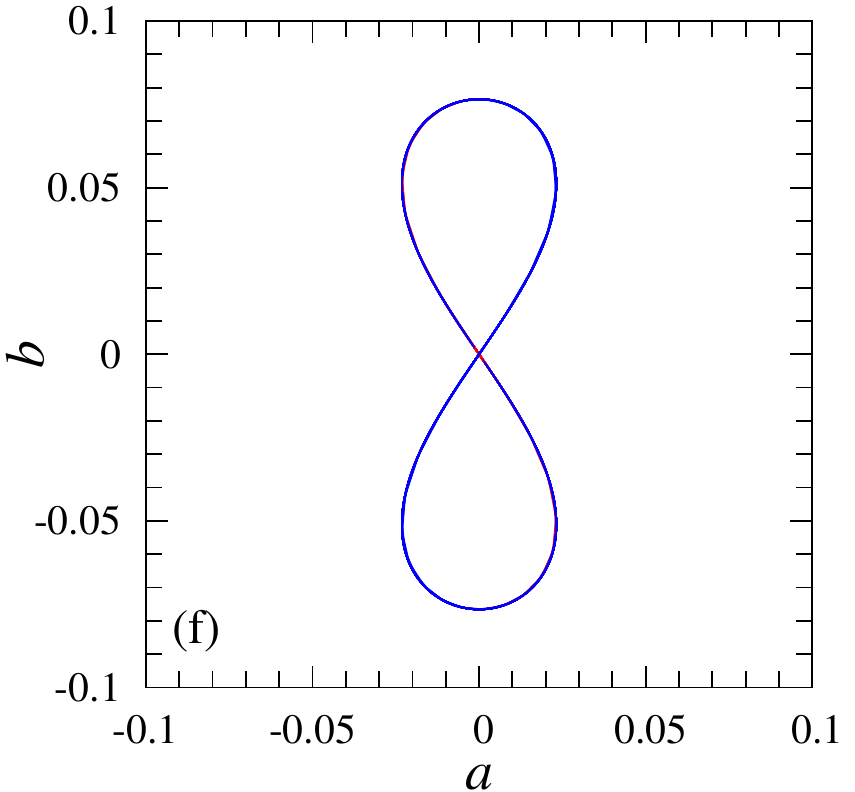}
\caption{Stable and unstable manifolds of the origin on the Poincar\'{e}
section $x=0\mod 2\protect\pi$ when $\protect\mu=0$ and $k=1$: (a) $\protect%
\omega=-1.118$, $\protect\varepsilon=0.045$; (b) $\protect\omega=1.118$, $%
\protect\varepsilon=0.045$; (c) $\protect\omega=-1.42$, $\protect\varepsilon%
=0.4$; (d) $\protect\omega=1.42$, $\protect\varepsilon=0.4$; (e) $\protect%
\omega=-1.9$, $\protect\varepsilon=0.6$; (f) $\protect\omega=1.9$, $\protect%
\varepsilon=0.6$. The red and blue curves represent the stable and unstable
manifolds, respectively.}
\label{f:im123}
\end{center}
\end{figure}

\begin{figure}[t]
\begin{center}
\scalebox{0.99}{
\begin{minipage}{.48\textwidth}
\begin{center}
\includegraphics[width=.85\textwidth]{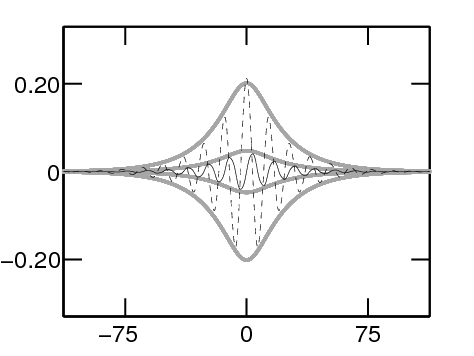} \\[-0.2ex]
(a) $\omega=-1.118$, $\epsilon=0.045$, $\mu=0$
\end{center}
\end{minipage}
\begin{minipage}{.48\textwidth}
\begin{center}
\includegraphics[width=.85\textwidth]{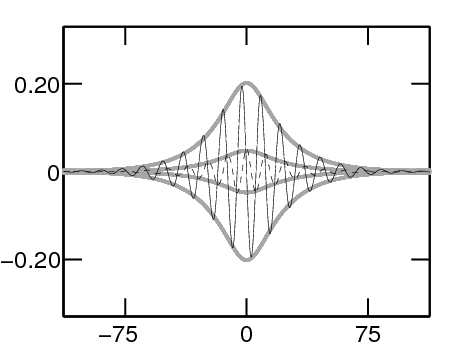} \\[-0.2ex]
(b) $\omega=1.118$, $\epsilon=0.045$, $\mu=0$
\end{center}
\end{minipage}
} \vspace*{2ex}
\par
\scalebox{0.99}{
\begin{minipage}{.48\textwidth}
\begin{center}
\includegraphics[width=.85\textwidth]{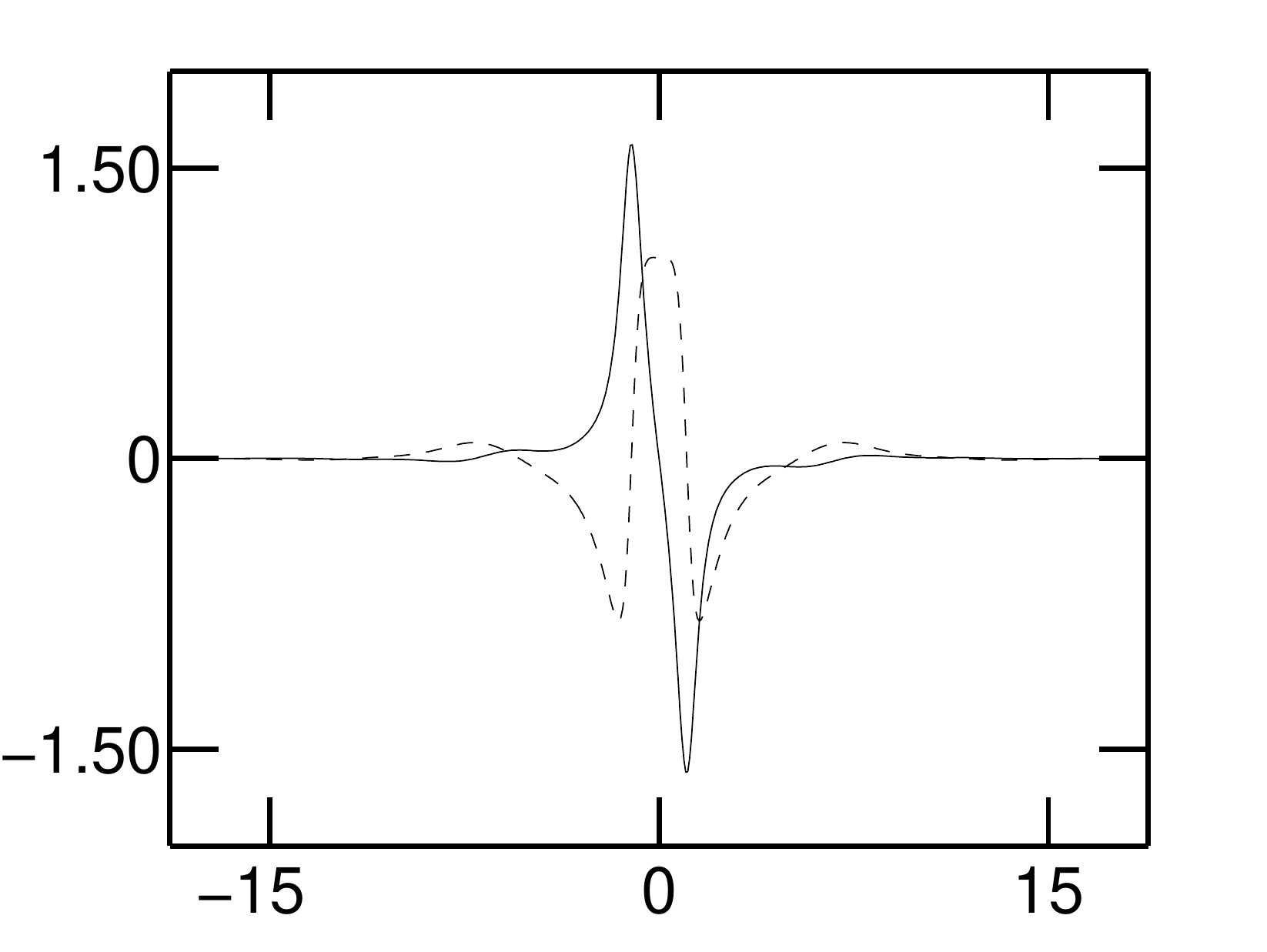} \\[-0.2ex]
(c) $\omega=-1.118$, $\epsilon=0.4$, $\mu=0$
\end{center}
\end{minipage}
\begin{minipage}{.48\textwidth}
\begin{center}
\includegraphics[width=.85\textwidth]{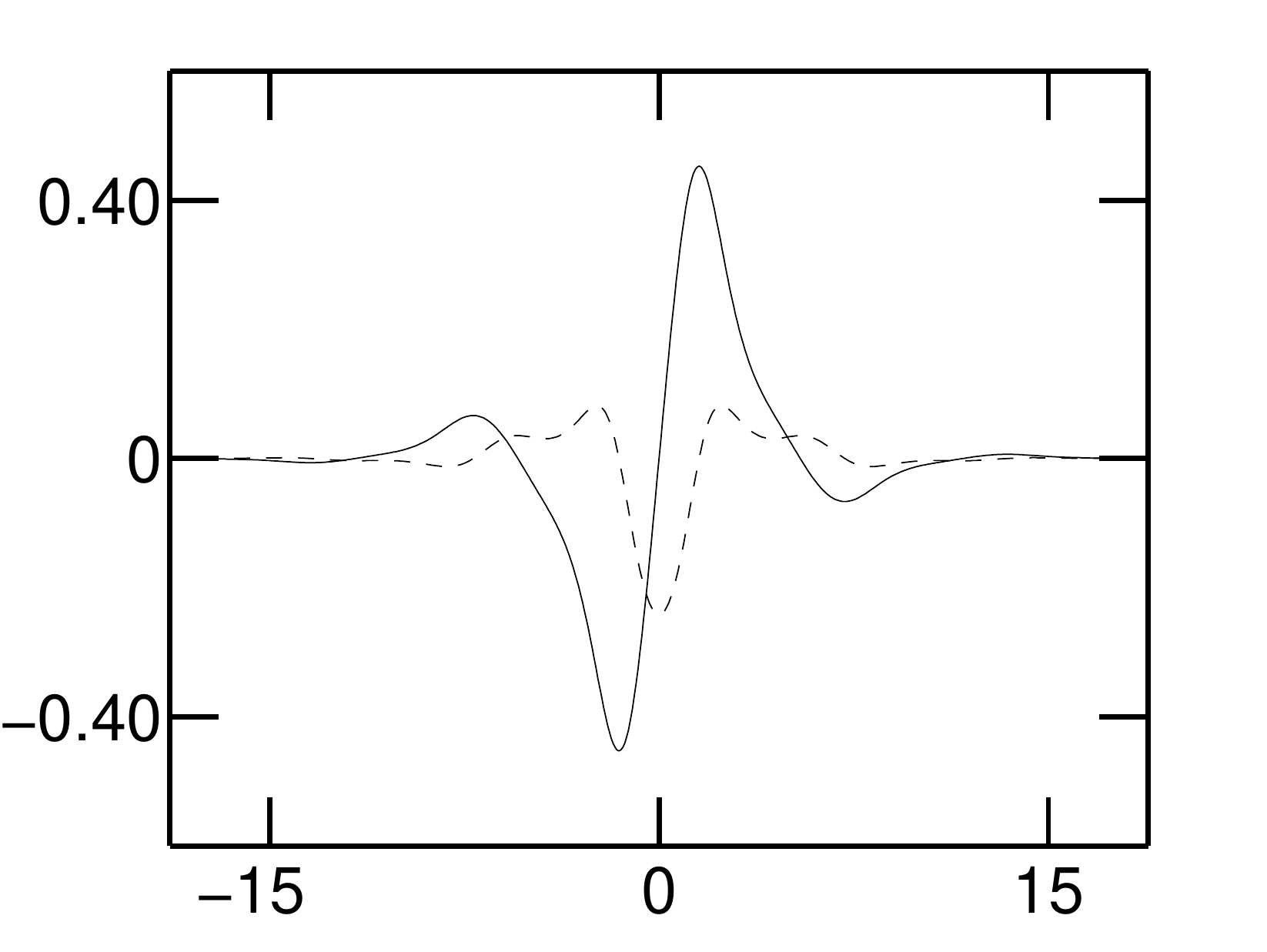} \\[-0.2ex]
(d) $\omega=1.118$, $\epsilon=0.4$, $\mu=0$
\end{center}
\end{minipage}
}
\end{center}
\caption{GSs with $V=-U^*$ of (\protect\ref{e:uvans}) in the gaps
$\mathbf{1^\pm}$, $\mathbf{2^\pm}$ and $\mathbf{3^\pm}$ for $k=1$,
symmetric with respect to $R_1$ or $R_2$. Gray curves in
the panels correspond to solutions of the associated averaged system, (%
\protect\ref{e:av1}), (\protect\ref{e:av2}) or (\protect\ref{e:av3}). }
\end{figure}

\addtocounter{figure}{-1}
\begin{figure}[t]
\begin{center}
\scalebox{0.99}{
\begin{minipage}{.48\textwidth}
\begin{center}
\includegraphics[width= .85\textwidth]{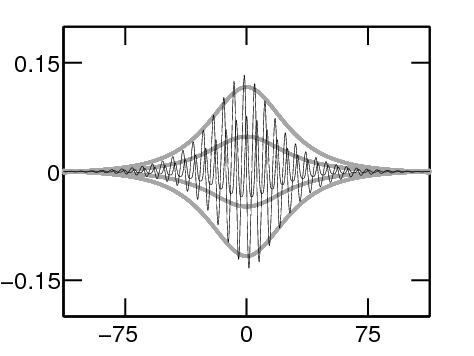} \\[-0.2ex]
(e) $\omega=-1.42$, $\epsilon=0.4$, $\mu=0$
\end{center}
\end{minipage}
\begin{minipage}{.48\textwidth}
\begin{center}
\includegraphics[width=.85\textwidth]{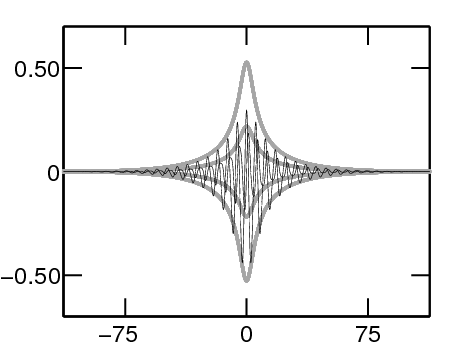} \\[-0.2ex]
(f) $\omega=1.42$, $\epsilon=0.4$, $\mu=0$
\end{center}
\end{minipage}
} \vspace*{2ex}
\par
\scalebox{0.99}{
\begin{minipage}{.48\textwidth}
\begin{center}
\includegraphics[width= .85\textwidth]{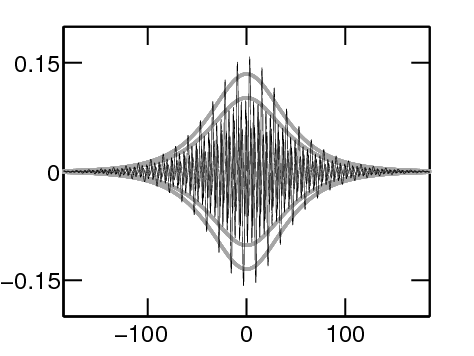} \\[-0.2ex]
(g) $\omega=-1.9$, $\epsilon=0.6$, $\mu=0$
\end{center}
\end{minipage}
\begin{minipage}{.48\textwidth}
\begin{center}
\includegraphics[width=.85\textwidth]{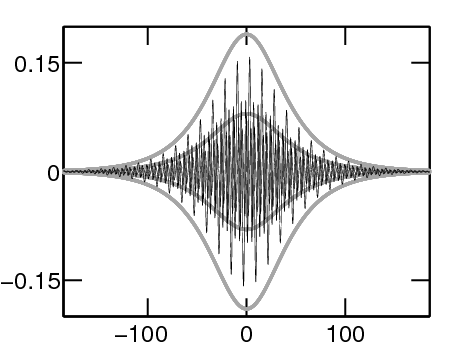} \\[-0.2ex]
(h) $\omega=1.9$, $\epsilon=0.6$, $\mu=0$
\end{center}
\end{minipage}
}
\caption{Continued.}
\label{f:big123}
\end{center}
\end{figure}

\begin{figure}[t]
\begin{minipage}{.48\textwidth}
\begin{center}
\includegraphics[width= .85\textwidth]{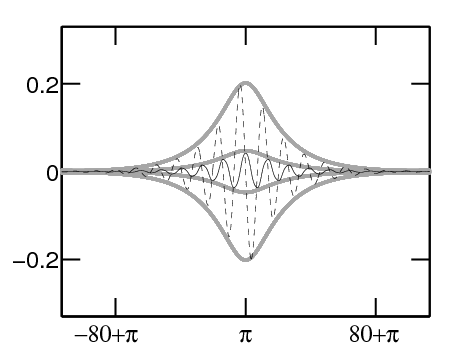} \\[1ex]
(a) $\omega=-1.118$, $\epsilon=0.045$, $\mu=0$
\end{center}
\end{minipage}
\begin{minipage}{.48\textwidth}
\begin{center}
\includegraphics[width=.85\textwidth]{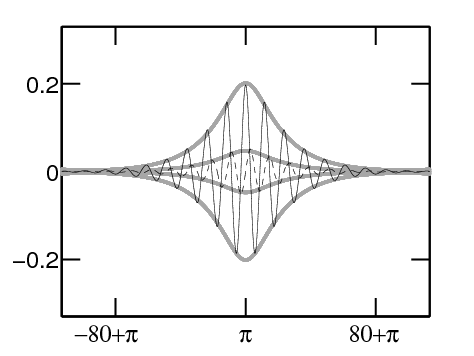} \\[1ex]
(b) $\omega=1.118$, $\epsilon=0.045$, $\mu=0$
\end{center}
\end{minipage}
\caption{GSs with $V=-U^*$ of (\protect\ref{e:uvans}) in the gaps $\mathbf{%
1^\pm}$ for $k=1$, symmetric with respect to $R_1^{\prime}$ or $R_2^{\prime}$%
. Gray curves in the panels correspond to solutions of the averaged system (%
\protect\ref{e:av1}).}
\end{figure}

\begin{figure}[t]
\begin{minipage}{.48\textwidth}
\begin{center}
\includegraphics[width=.85\textwidth]{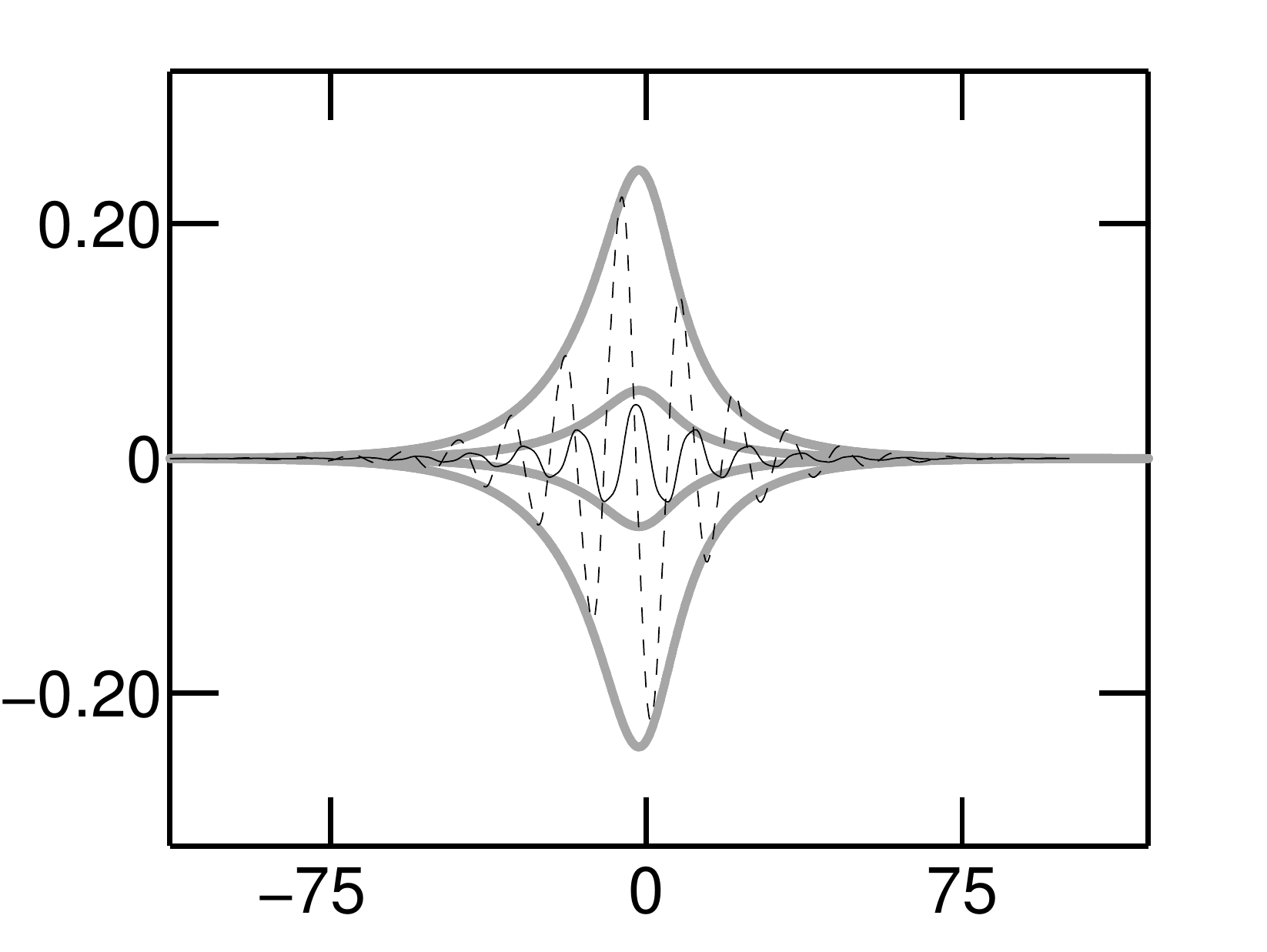} \\[1ex]
(a) $\omega=-1.118$, $\epsilon=0.045$, $\mu=0.045$, $\delta=\pi/2$
\end{center}
\end{minipage}
\begin{minipage}{.48\textwidth}
\begin{center}
\includegraphics[width=.85\textwidth]{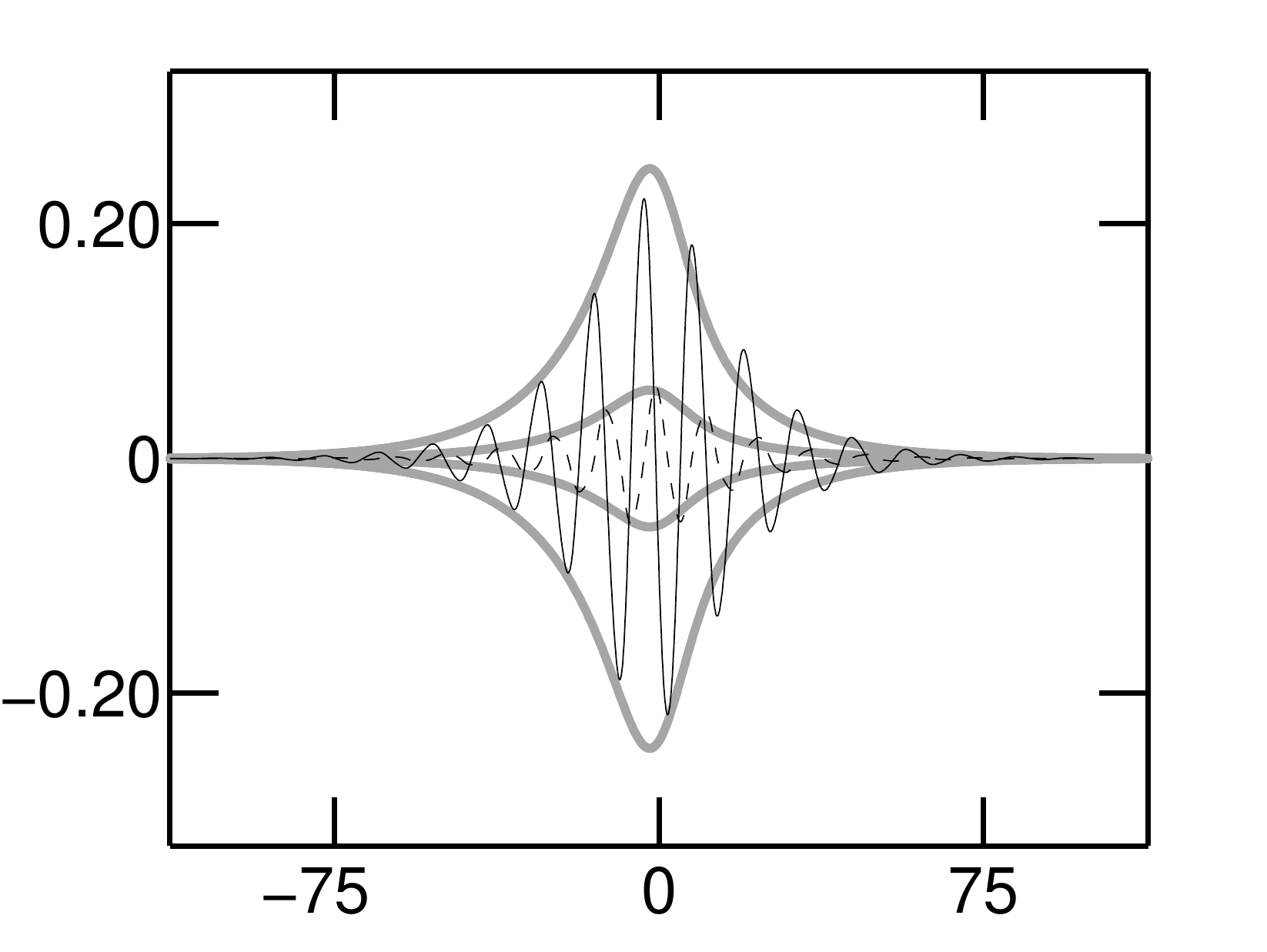} \\[1ex]
(b) $\omega=1.118$, $\epsilon=0.045$, $\mu=0.045$, $\delta=\pi/2$
\end{center}
\end{minipage}
\vspace*{2ex}
\caption{Asymmetric GSs with $V=-U^*$ of (\protect\ref{e:uvans}) in the gaps
$\mathbf{1^\pm}$ for $k=1$. Gray curves in the panels correspond to
solutions of the averaged system (\protect\ref{e:av1}). }
\label{f:big1}
\end{figure}

\begin{figure}[t]
\begin{minipage}{.48\textwidth}
\begin{center}
\includegraphics[width=.85\textwidth]{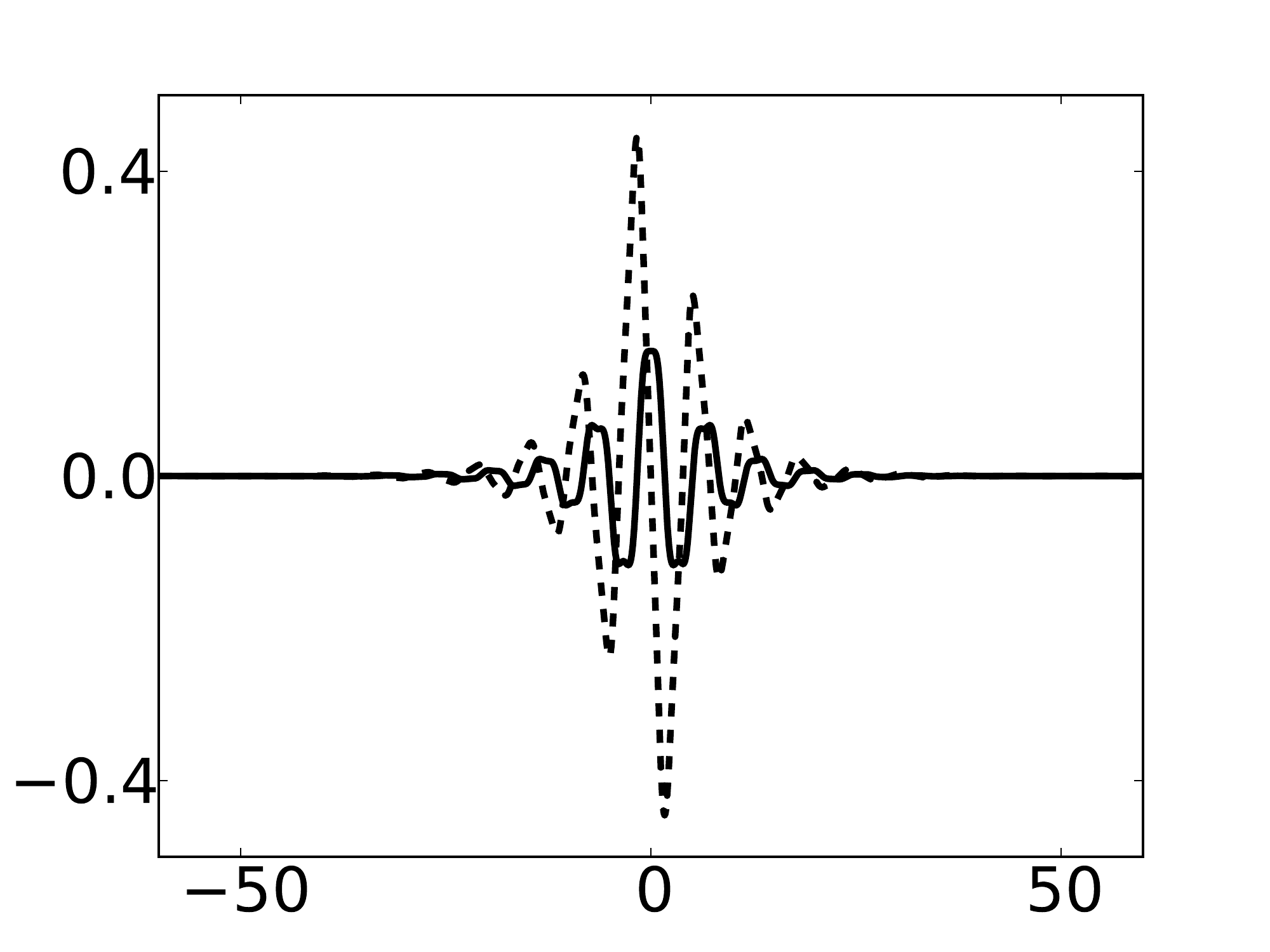} \\[1ex]
(a) $\epsilon=0.2$
\end{center}
\end{minipage}
\begin{minipage}{.48\textwidth}
\begin{center}
\includegraphics[width=.85\textwidth]{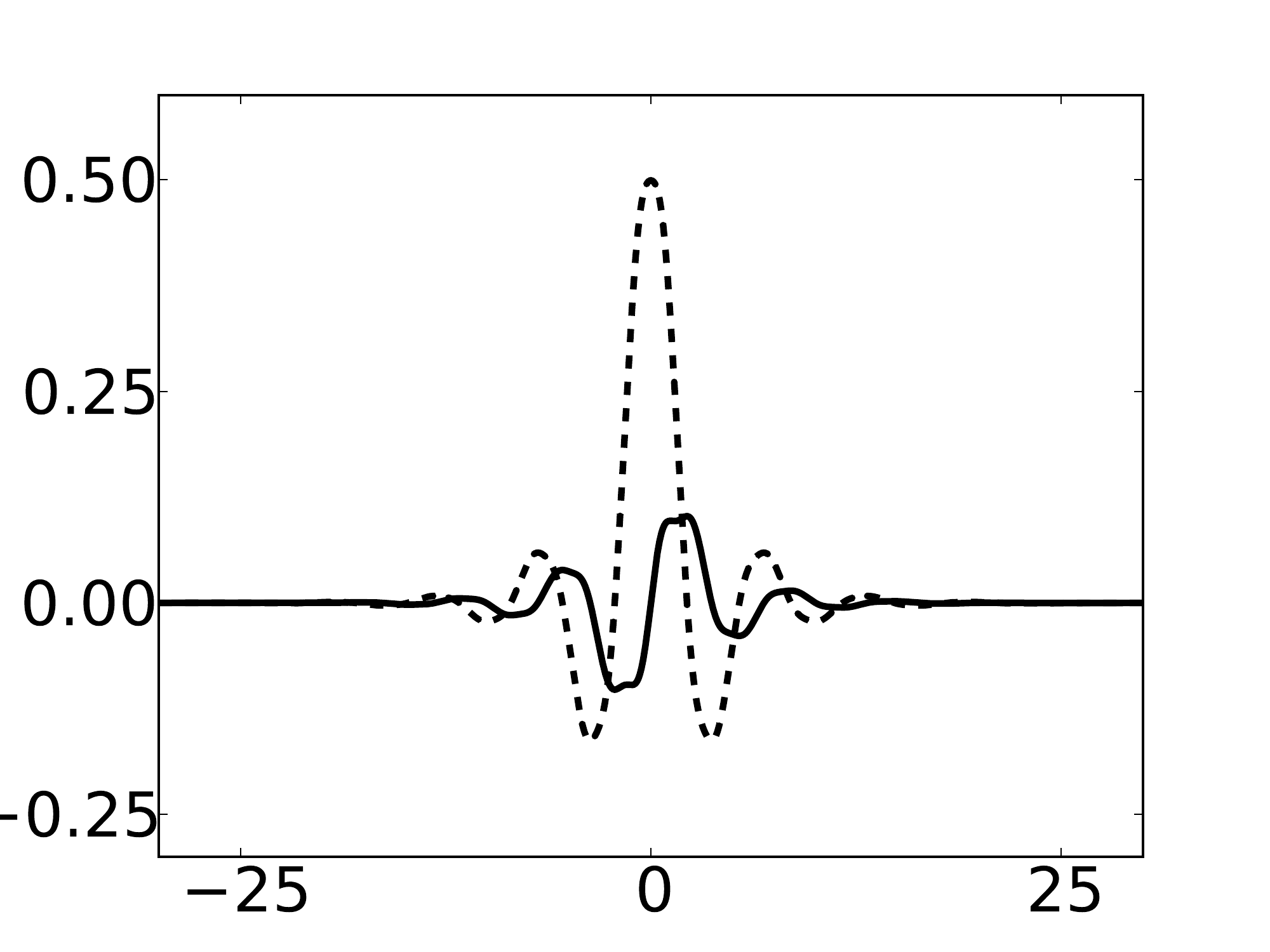} \\[1ex]
(b) $\epsilon=0.8$
\end{center}
\end{minipage}
\vspace*{2ex}
\caption{Symmetric GSs with $V=-U^*$ of (\protect\ref{e:uvans}) in the gaps $%
\mathbf{1^\pm}$ for $\protect\omega=-1.4$, $\protect\mu=0.5$, $\protect\delta%
=0$ and $k=2$. }
\label{f:big1a}
\end{figure}

We computed the stable and unstable manifolds of the origin and homoclinic
orbits in (\ref{e:ab}), i.e., GSs with $V=-U^*$ in (\ref{e:uv}), using the
computer package, \texttt{AUTO} with \texttt{HomMap}. In our computations we
found GSs due to the transverse intersection of the stable and unstable
manifolds, thus confirming the analytical results. For several parameter
values, the stable and unstable manifolds for the Poincar\'{e} section $x=0%
\mod 2\pi$ are shown in Fig.~\ref{f:im123} and GSs in (\ref{e:uv}) are
plotted in Figs.~\ref{f:big123}-\ref{f:big1}. In addition to the solutions
of (\ref{e:ab}), the figures include homoclinic solutions of the averaged
systems (\ref{e:av1}), (\ref{e:av2}) and (\ref{e:av3}), rescaled according
to the transformation~(\ref{e:trans}). As can be seen, they provide a good
match to the envelope of the solutions to (\ref{e:ab}). In all figures
presented, the parameter $k$ has been set to $k=1$.

In closing this section, we return to the parameter values of Fig.~\ref%
{f:linear}(d). In this case a new bandgap opens up for small $\varepsilon$,
but closes at $\varepsilon=1/\sqrt{8}$. See also formula~\eqref{e:closing}.
Our computations reveal that GSs exist in the gaps both above and below this
parameter value. In Fig.~\ref{f:big1a} we show solution profiles at $%
\varepsilon=0.2$ and $\varepsilon=0.8$.

\section{Bifurcation of solitons}

\label{s:bif}

We further discuss the behavior of GSs under variation of the system
parameters. When the parameter values are varied, homoclinic tangencies,
i.e., homoclinic bifurcations, can occur, leading to the creation of
additional GSs and bifurcations of GSs. We have found such bifurcations in
all gap regions including the central one.

\begin{figure}[t]
\begin{center}
\includegraphics[width=0.46\textwidth]{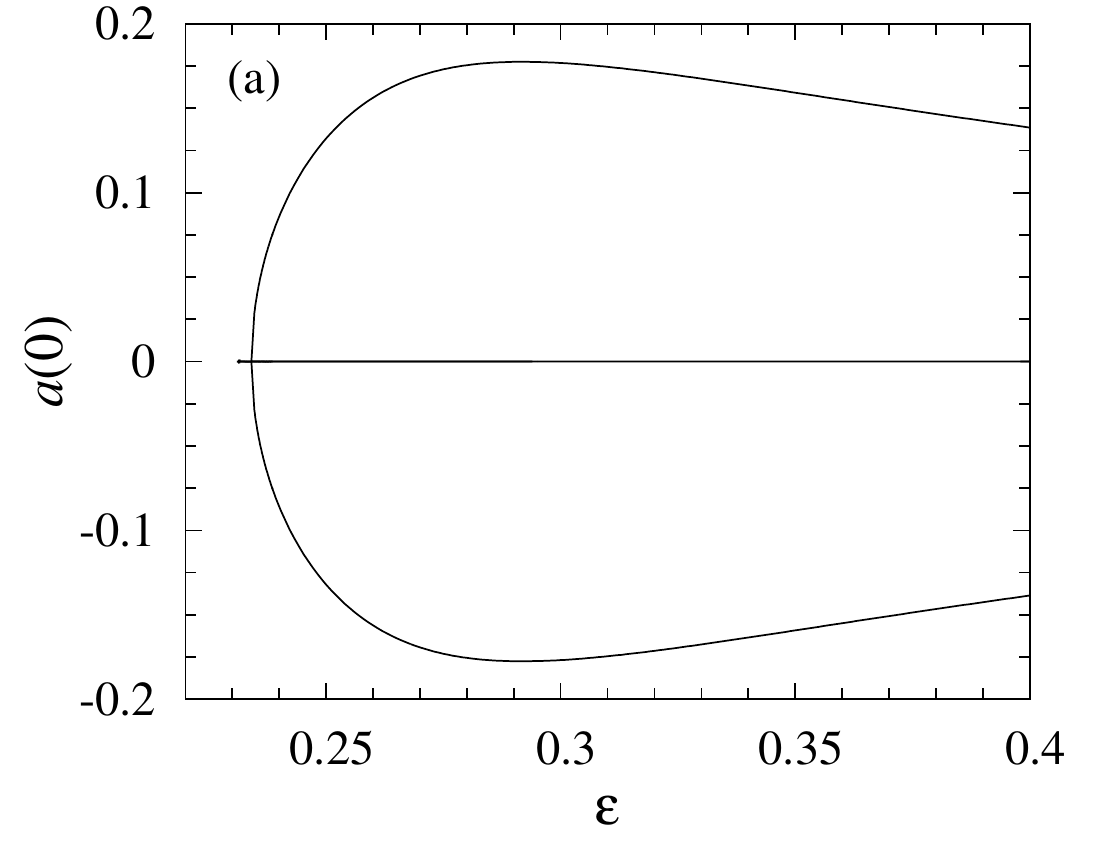}\quad %
\includegraphics[width=0.46\textwidth]{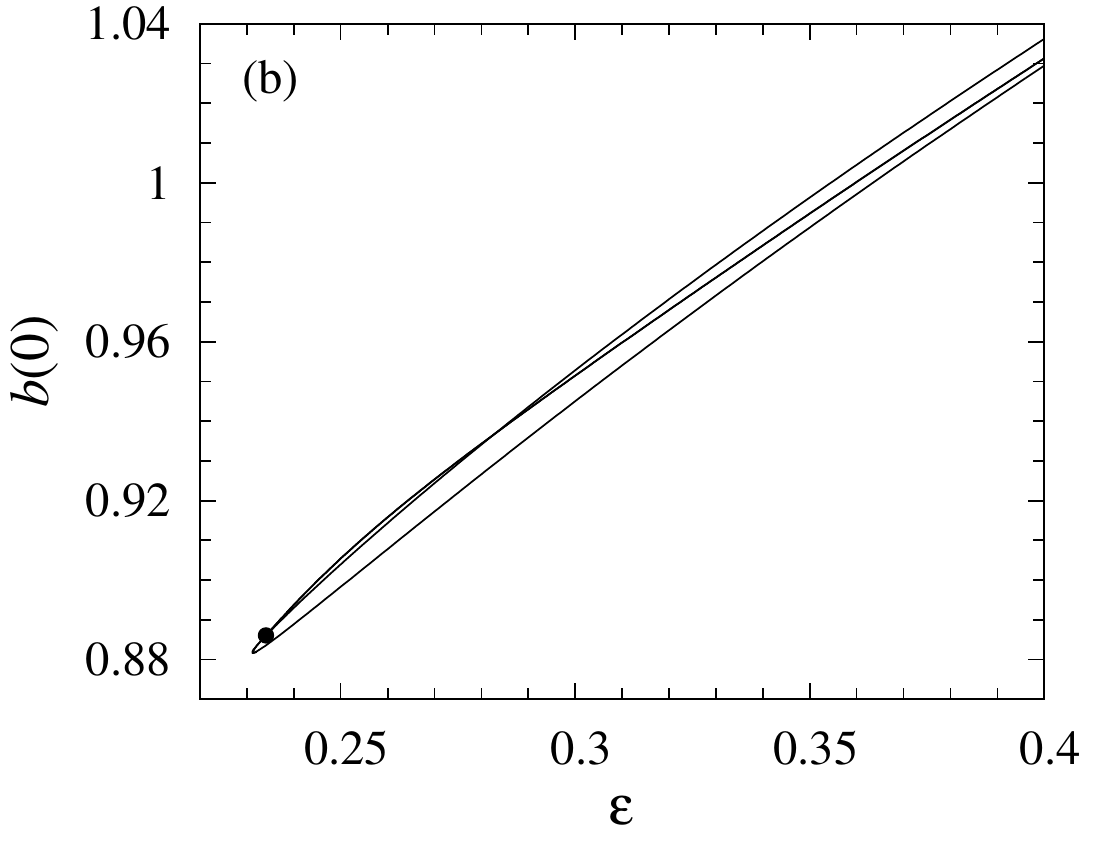}\\[2ex]
\includegraphics[width=0.46\textwidth]{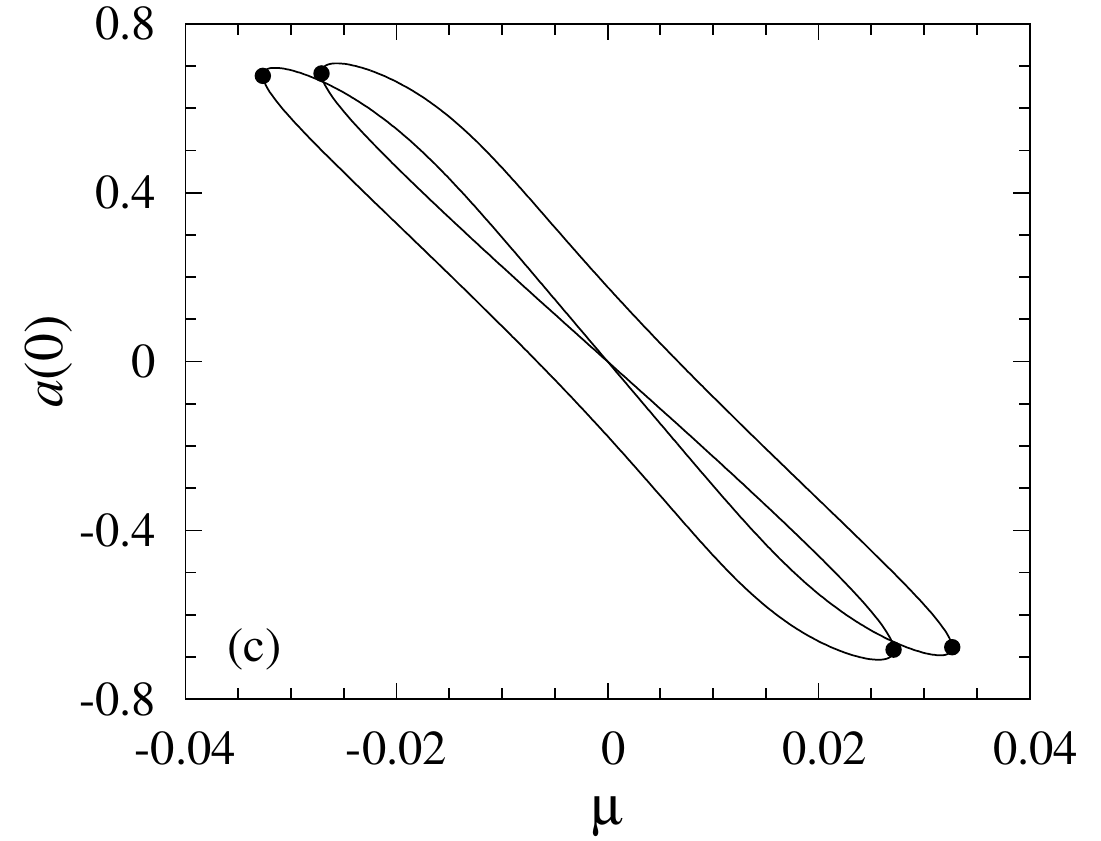}\quad %
\includegraphics[width=0.46\textwidth]{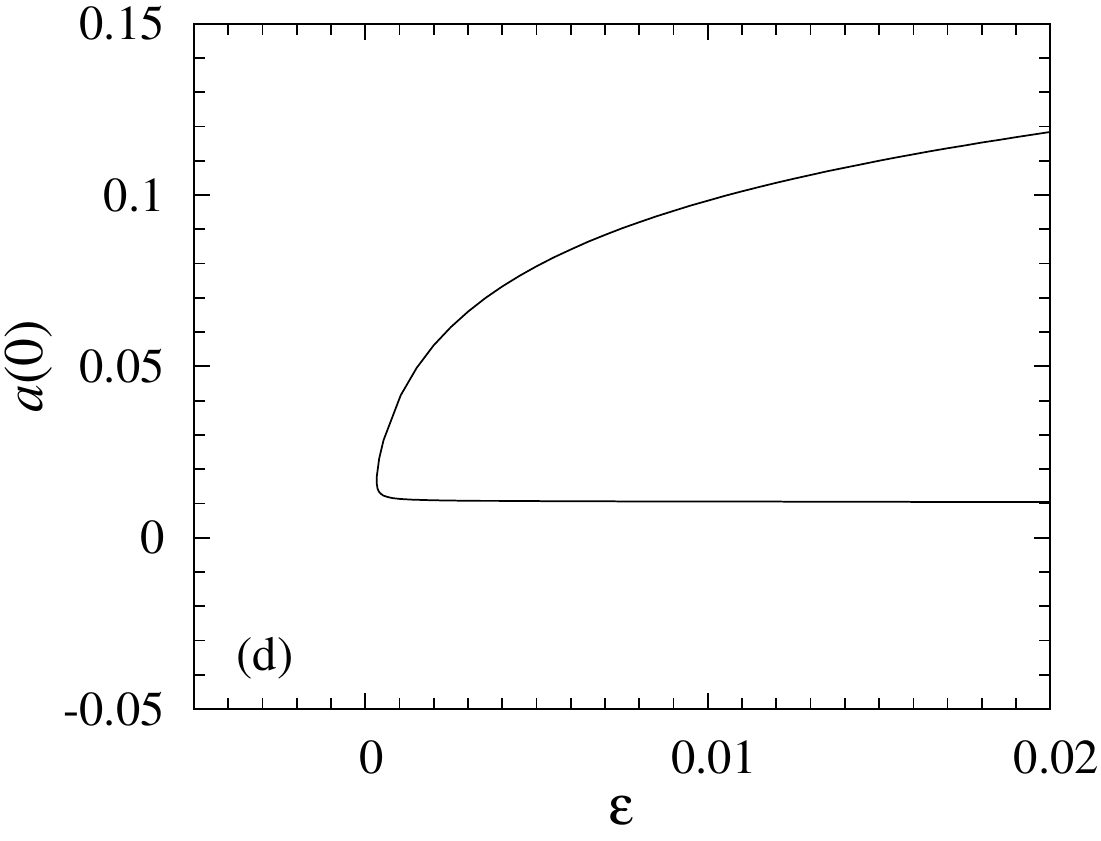}\\[0pt]
\caption{Bifurcation diagrams of GSs in Eqs. (\protect\ref{e:uvans}) in the
gaps $\mathbf{1^-}$ and $\mathbf{0}$ for $k=1$ and $\protect\delta=\protect%
\pi/2$: (a) and (b) $\protect\omega=-1.118$, $\protect\mu=0$; (c) $\protect%
\omega=-1.118$, $\protect\varepsilon=0.3$; (d) $\protect\omega=0.5$, $%
\protect\mu=0$. In plates~(b) and (c), the point ``$\bullet$'' represents
pitchfork and saddle-node bifurcations, respectively}
\label{f:bd}
\end{center}
\end{figure}

As an example, we plot bifurcation diagrams of GSs for Eqs. (\ref{e:uvans})
in the gaps $\mathbf{1^-}$ and $\mathbf{0}$ when $\omega=-1.118$ or 0.5 and $%
\delta=\pi/2$, in Fig.~\ref{f:bd}. Here $(a(0),b(0))$ represents the point
at which each GS crosses the section $\Sigma=\{x=0\mod 2\pi\}$. In the
diagrams we observe saddle-node and pitchfork bifurcations of GSs. The
occurrence of these bifurcations is explained from the behavior of the
Poincar\'{e} map of (\ref{e:ab}) for the section $\Sigma$, as follows.

\begin{figure}[t]
\begin{center}
\includegraphics[scale=0.7]{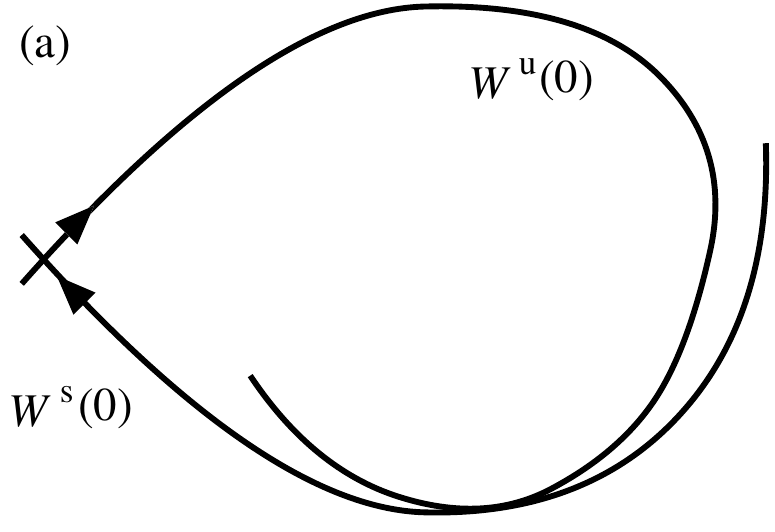}\qquad %
\includegraphics[scale=0.7]{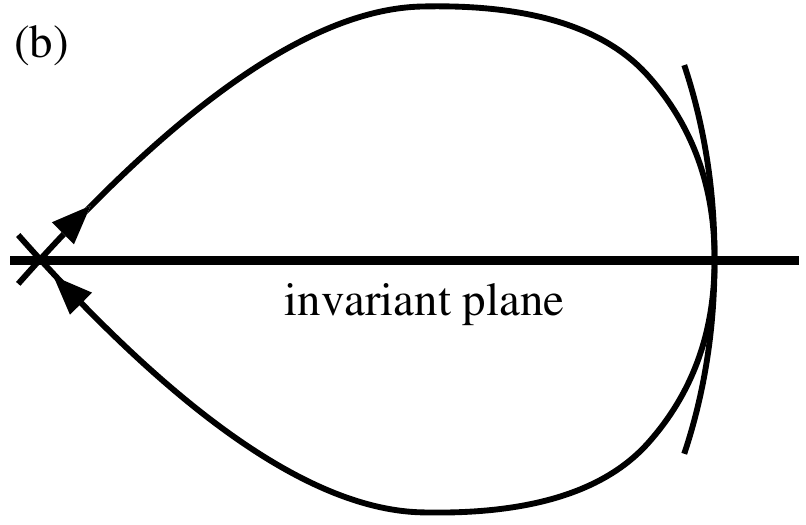}\\[2ex]
\includegraphics[scale=0.7]{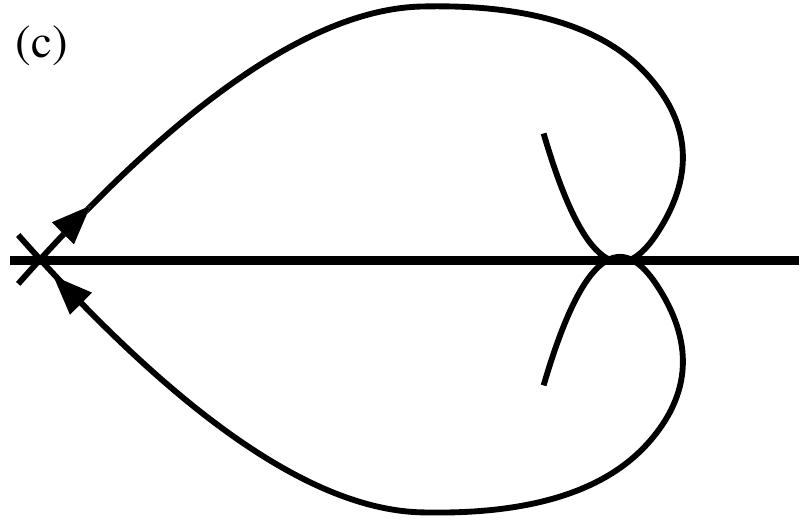}
\caption{Homoclinic bifurcations in the Poincar\'{e} map. Panel (a) depicts
the situation that leads to a saddle-node bifurcation of asymmetric GSs.
Panels~(b) and (c), respectively, explain mechanisms for pitchfork and
saddle-node bifurcations of symmetric GSs.}
\label{f:hbmap}
\end{center}
\end{figure}

Let us first assume that $\mu \neq 0$, such that the Poincar\'e map is not
reversible. In this case, a tangency between the stable and unstable
manifolds as shown in Fig.~\ref{f:hbmap}(a) leads to a saddle-node
bifurcation, in which two new GSs are created. See Fig.~\ref{f:bd}(c) for a
numerical evidence.

Now suppose that $\mu=0$ and let us choose a cross-section, such that the
Poincar\'{e} map is reversible with respect to $R_1$ and $R_2$. We can
distinguish two different bifurcation scenarios.

If the stable and unstable manifolds have a cubic tangency on the invariant
plane of $R_1$ or $R_2$, but intersect this plane transversely as shown in
Fig.~\ref{f:hbmap}(b), then a pitchfork bifurcation of homoclinic orbits
takes place. When the parameters are varied, a pair of homoclinic orbits
appear, while the original homoclinic orbit still exists due to the
persistence of symmetric orbits. See Figs.~\ref{f:bd}(a) and (b). The
original GS is symmetric, but the two new orbits are asymmetric.

On the other hand, if the stable and unstable manifolds are tangent to the
invariant plane at their quadratic tangency as shown in Fig.~\ref{f:hbmap}%
(c), then a saddle-node bifurcation of symmetric GSs occurs. See Fig.~\ref%
{f:bd}(b). Note that a saddle-node bifurcation of asymmetric GSs can occur
even though $\mu=0$.

\begin{figure}[t]
\begin{center}
\includegraphics[scale=0.65]{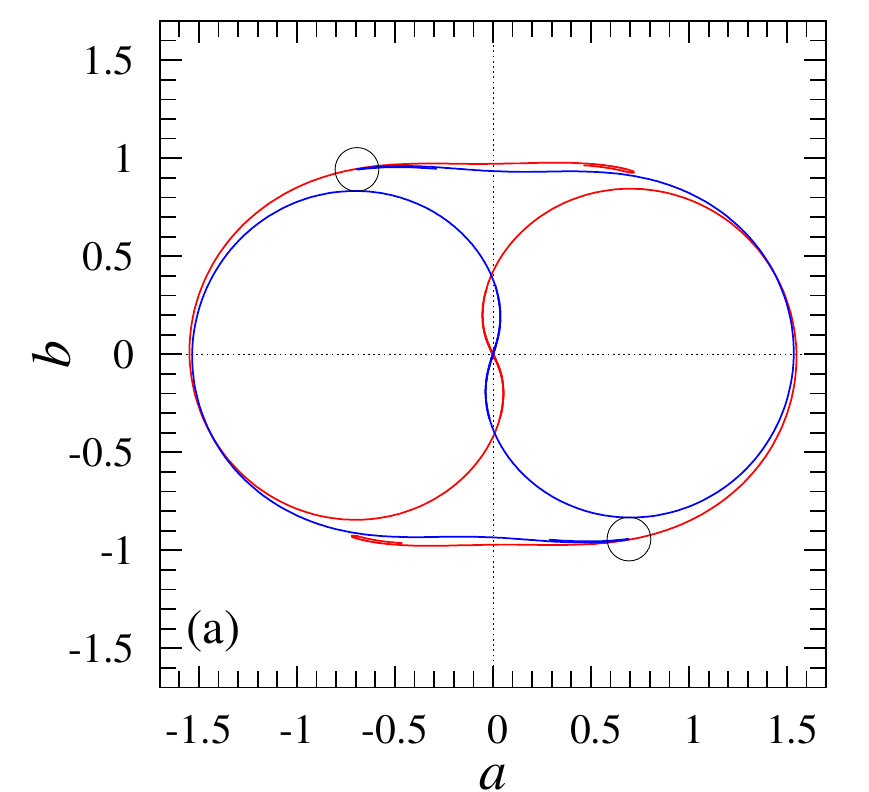} \qquad %
\includegraphics[scale=0.65]{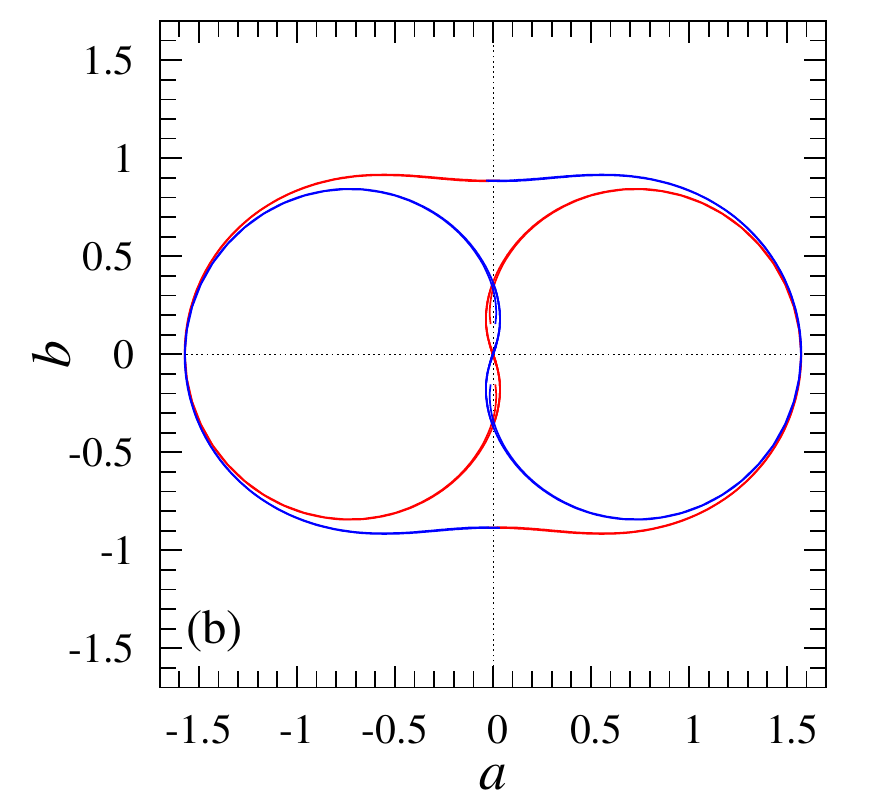}\\[3ex]
\includegraphics[scale=0.65]{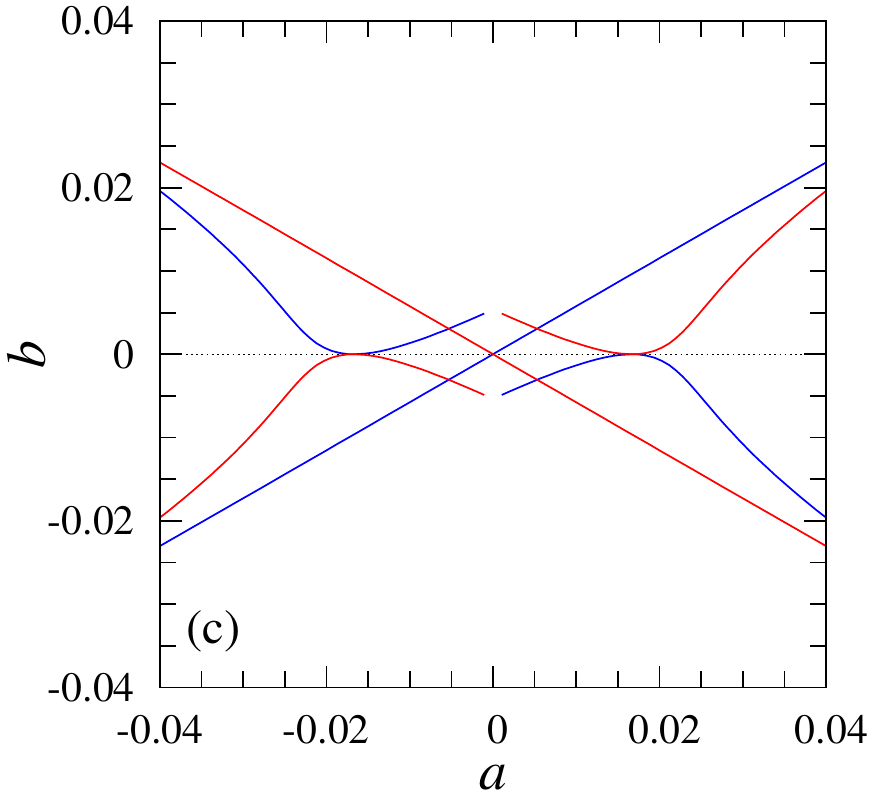}
\caption{Stable and unstable manifolds of the origin on the Poincar\'{e}
section $x=0\mod 2\protect\pi$ when $k=1$ and $\protect\delta=\protect\pi/2$%
: (a) $\protect\omega=-1.118$, $\protect\varepsilon=0.3$, $\protect\mu%
=0.0326551$; (b) $\protect\omega=-1.118$, $\protect\varepsilon=0.234076$, $%
\protect\mu=0$; (c) $\protect\omega=0.5$, $\protect\varepsilon=3.35338\times
10^{-4}$, $\protect\mu=0$. The red and blue curves represent the stable and
unstable manifolds, respectively. In plate~(a) the invariant manifolds have
quadratic tangencies inside the circles.}
\label{f:im}
\end{center}
\end{figure}

We illustrate these general ideas with computations for equation \eqref{e:ab}%
. Figure~\ref{f:im} shows the stable and unstable manifolds of the
Poincar\'e map of \eqref{e:ab} on the section $\Sigma$ for the parameter
values when saddle-node or pitchfork bifurcations occur in Figs.~\ref{f:bd}.

Figure~\ref{f:im}(a) corresponds to Fig.~\ref{f:hbmap}(a) and depicts the
situation when the map is not reversible ($\mu>0$) and the stable and
unstable manifolds have a quadratic tangency. Similarly, in Figs.~\ref{f:im}%
(b) and (c) the Poincar\'e map is reversible, and the manifolds have cubic
and quadratic tangencies on the invariant plane $b=0$ as in the sketches in
Figs.~\ref{f:hbmap}(b) and (c).

\begin{figure}[t]
\begin{minipage}{.32\textwidth}
\begin{center}
\includegraphics[width=\textwidth]{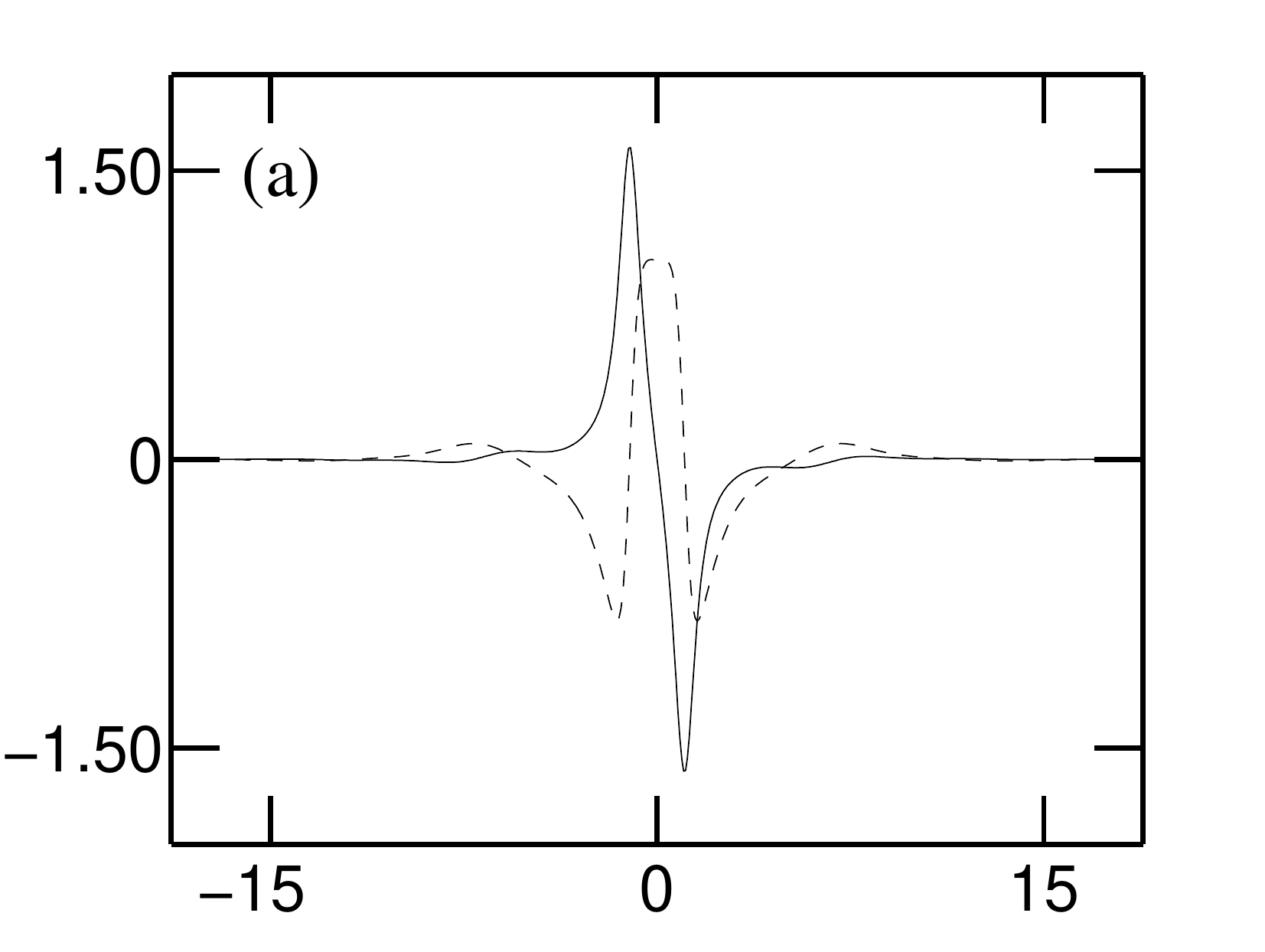}
\end{center}
\end{minipage}
\begin{minipage}{.32\textwidth}
\begin{center}
\includegraphics[width=\textwidth]{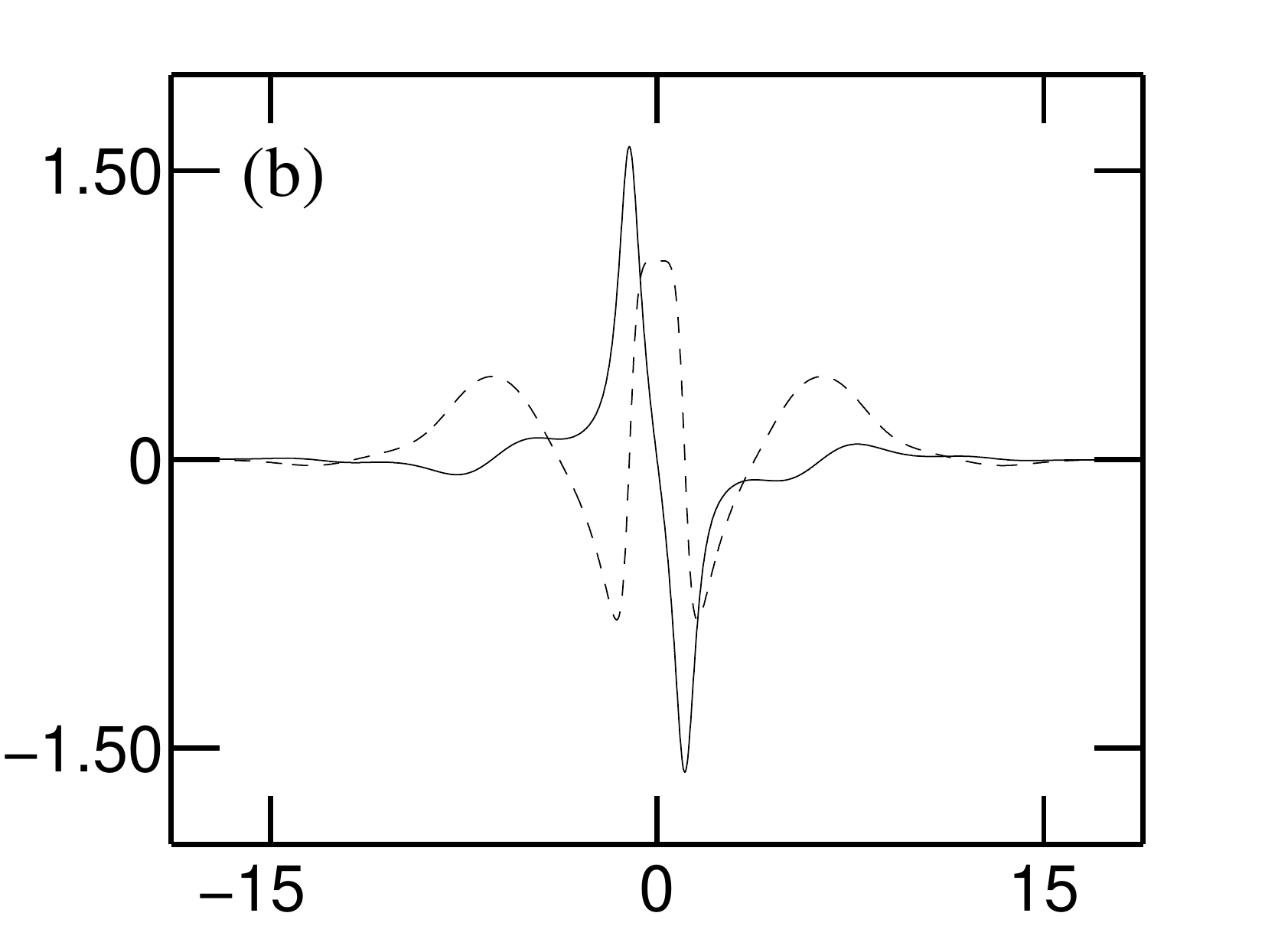}
\end{center}
\end{minipage}
\begin{minipage}{.32\textwidth}
\begin{center}
\includegraphics[width=\textwidth]{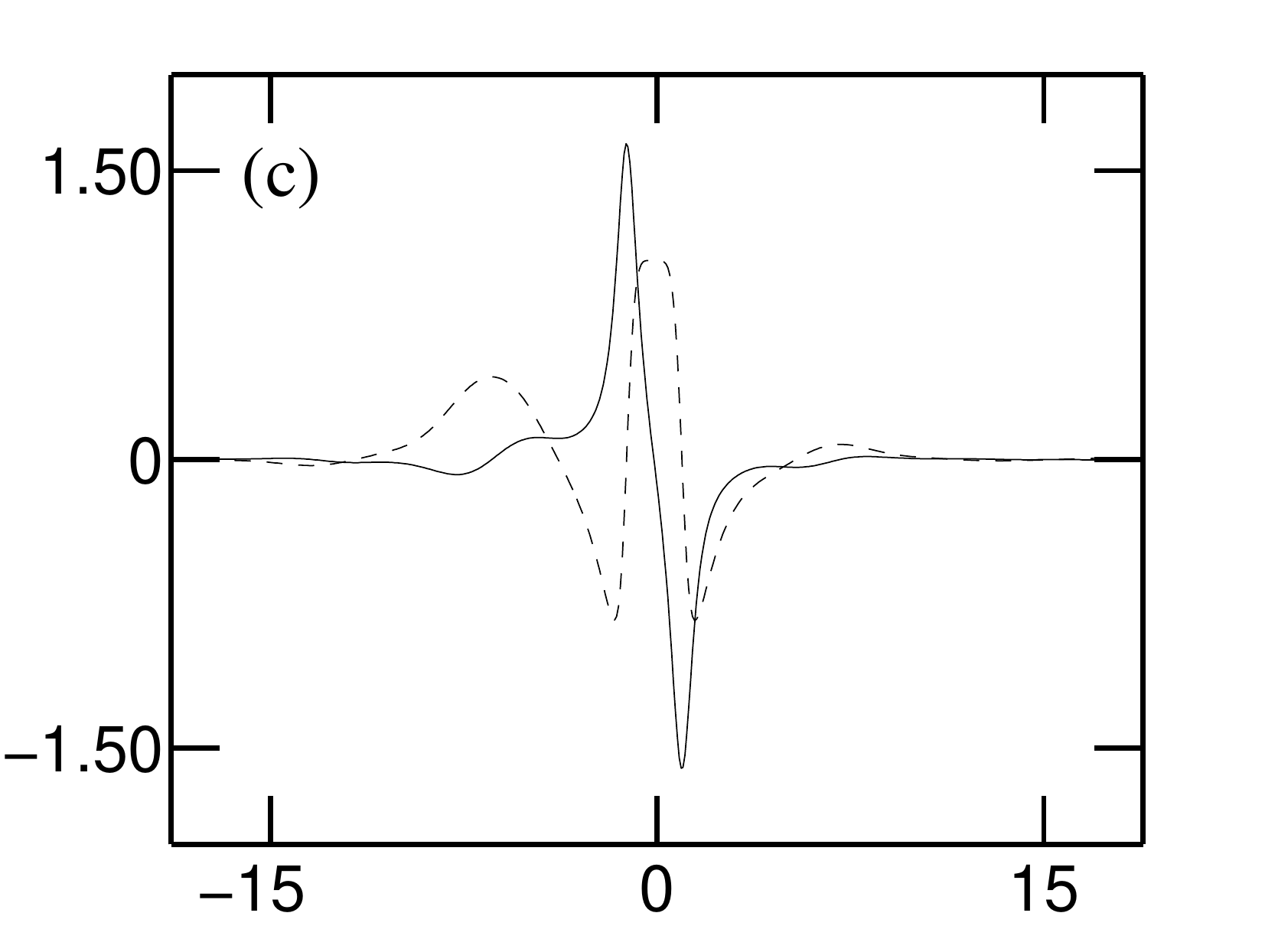}
\end{center}
\end{minipage}
\caption{GSs with $V=-U^*$ of (\protect\ref{e:uvans}) born at the
bifurcations of Fig.~\protect\ref{f:bd}(a)-(c) for $\protect\varepsilon=0.4$%
, $\protect\omega=-1.118$, $k=1$ and $\protect\mu=0$. The GSs in plates~(a)
and (b) are symmetric and born at the saddle-node bifurcation while the GSs
in plate~(c) is asymmetric and born at the pitchfork bifurcation. }
\label{f:newgs-1}
\end{figure}

\begin{figure}[t]
\begin{center}
\begin{minipage}{.32\textwidth}
\begin{center}
\includegraphics[width=\textwidth]{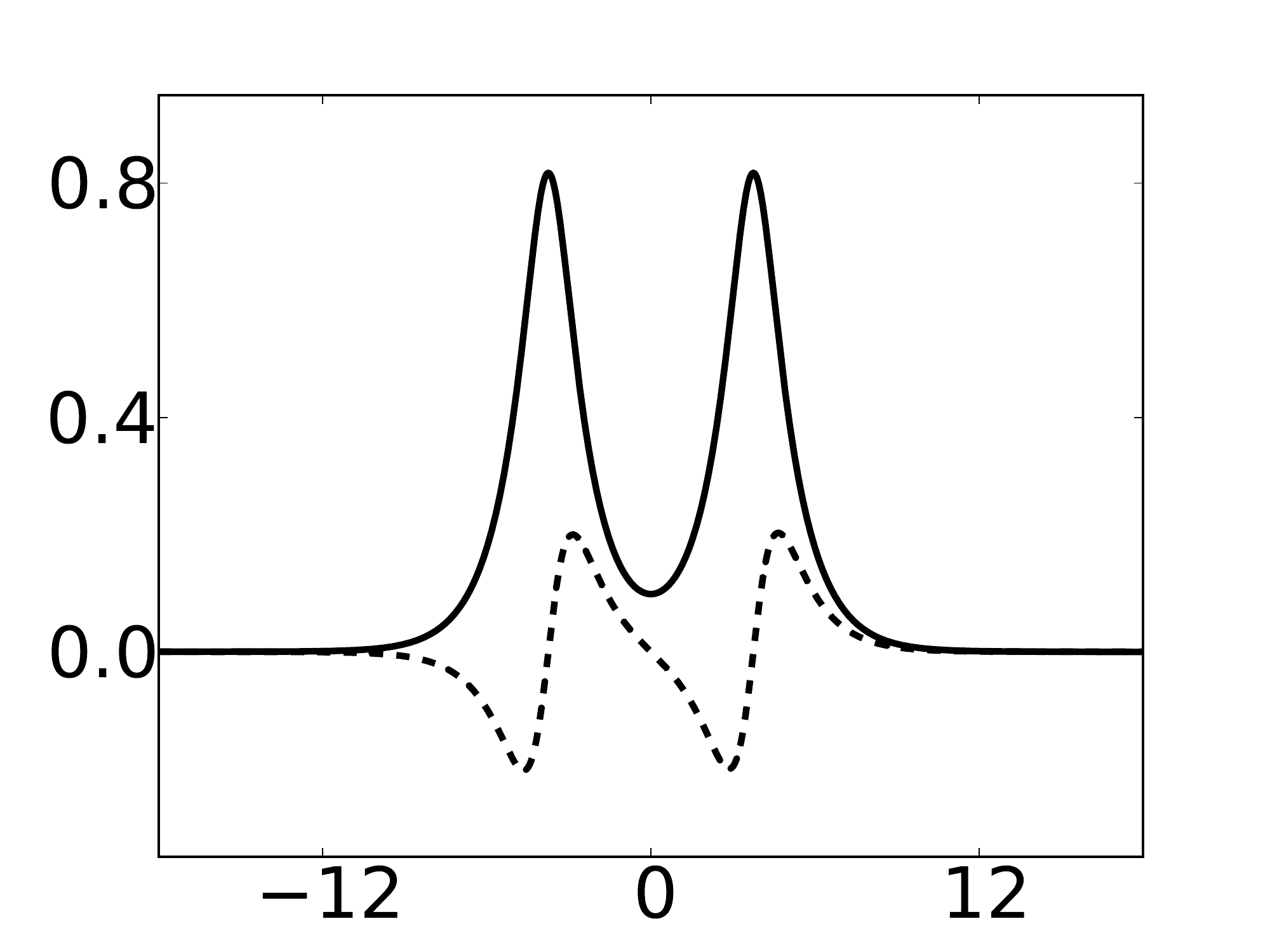}
\end{center}
\end{minipage}
\begin{minipage}{.32\textwidth}
\begin{center}
\includegraphics[width=\textwidth]{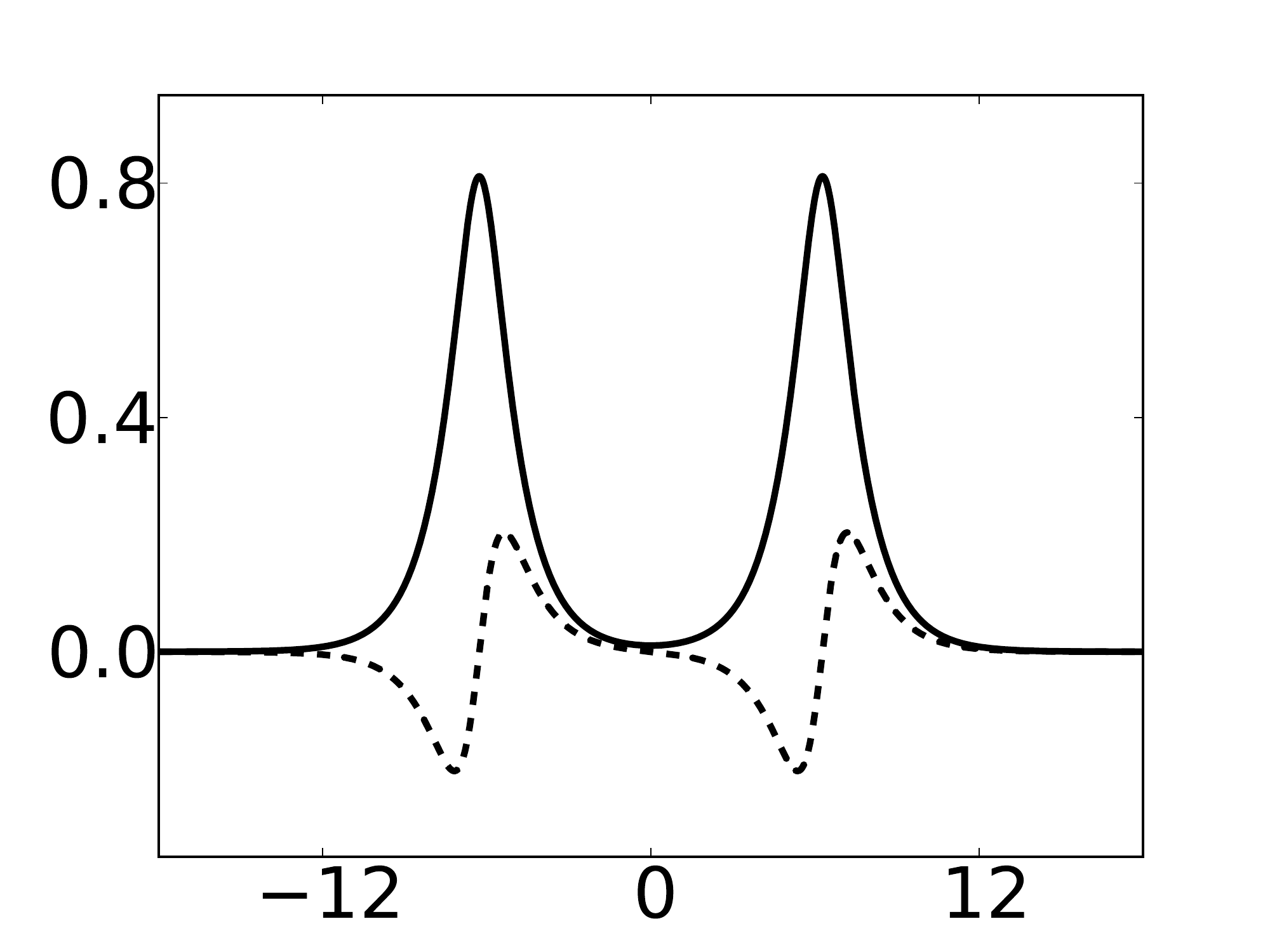}
\end{center}
\end{minipage}
\caption{Symmetric, two-pulse GSs with $V=-U^*$ of (\protect\ref{e:uvans})
born at the saddle-node bifurcation of Fig.~\protect\ref{f:bd}(d) for $%
\protect\varepsilon=0.01$, $\protect\omega=0.5$, $\protect\mu=0$ and $k=1$. }
\label{f:newgs0}
\end{center}
\end{figure}

Figures~\ref{f:newgs-1} and \ref{f:newgs0} show the profiles of GSs born at
the bifurcations in Figs.~\ref{f:bd}(a)-(c) and (d), respectively. In
particular, the GSs in Fig.~\ref{f:newgs0} are bound states with two humps.
Standard results about transverse homoclinic orbits explain the existence of
multi-pulse GSs, see \cite{Wi:90}, and we also find such GSs in the other
gap regions.

We note that bifurcations of symmetric homoclinic orbits in reversible
systems that are \emph{autonomous} have been discussed in \cite%
{BuChTo:96,Kn:97}. Our analysis of the Poincar\'e map also yields comparable
bifurcation results in systems with periodic forcing like Eqs.~(\ref{e:ab}).

\section{Conclusions}

In this paper we have studied the existence of GSs (gap solitons) in a model
of the BG (Bragg grating) subjected to periodic modulation along the fiber,
thus creating the supergrating. The corresponding mathematical model amounts
to a parametrically forced extension of the standard BG model with the
additional symmetry-breaking forcing term.

As was known previously \cite{BrSt:97}, the supergrating opens up a set of
new bandgaps. Using the Melnikov method and averaging techniques, we have
demonstrated that families of GSs exist robustly in the new gaps. We have
also analyzed bifurcations of GSs and showed that, in addition to the
fundamental GSs, the model also supports families of bound states of the
solitons.

Many of the GSs found in the model are completely stable, as was verified by
means of direct simulations for some of them \cite{YaMe:05}. The present
paper extends the results of \cite{YaMe:05} and gives detailed mathematical
proofs for them, although we did not analyze the stability problem here in
detail. In particular, we have discussed the effects of the
symmetry-breaking forcing term and demonstrated that it can lead to the
closing and re-opening of the bandgaps. In the presence of the new term, GSs
exist robustly (as generic solutions) in all the bandgaps, but, as the
symmetry of the equation is broken, they undergo many bifurcations.

\section*{Acknowledgments}

This research was initiated and many important parts were carried out when
the authors worked with Alan Champneys or visited him at the University of
Bristol. We thank him for discussions and comments as well as his
hospitality and support. K.Y. acknowledges support from the Japan Society
for the Promotion of Science, Grant-in-Aid for Scientific Research (C)
Nos.~18560056, 21540124 and 22540180.

\appendix

\section{Derivations of equations
 (\protect\ref{e:int1}) and (\protect\ref%
{e:int2})}

\label{a:cal}
\setcounter{figure}{0}

\begin{figure}[t]
\begin{center}
\includegraphics[scale=1]{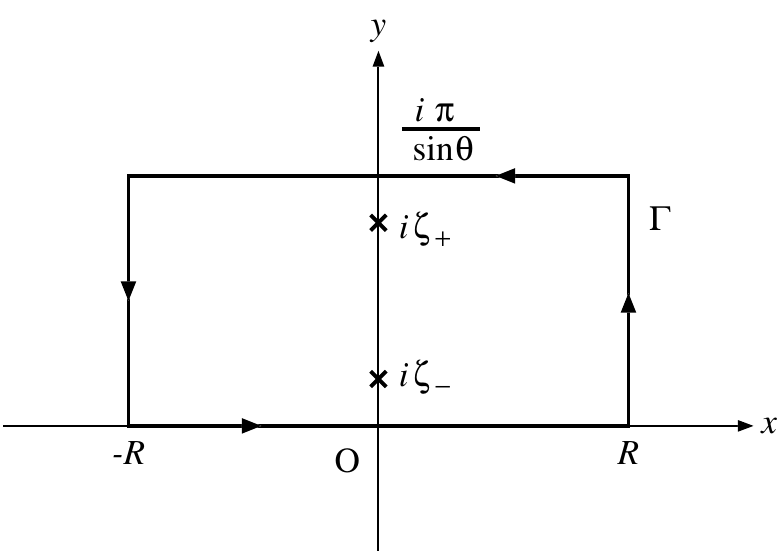}
\caption{The rectangle contour $\Gamma$.}
\label{f:a1}
\end{center}
\end{figure}

Let
\[
f_j(x)=\frac{\sinh(2 x\sin\theta)}{[\cosh(2x\sin\theta)+\cos\theta]^j}
\]
and let
\begin{equation}
I_j=\int_{-\infty}^\infty f_j(x)\sin kx\,d x  \label{e:Ij}
\end{equation}
for $j=2,3$. To estimate (\ref{e:Ij}), we consider the complex integral
\[
\oint_\Gamma f_j(z)\,e^{ikz}\,d z,
\]
where $\Gamma$ is a rectangle contour, $-R\le x\le R$ and $0\le
y\le\pi/\sin\theta$, depicted in Fig.~\ref{f:a1}, with $R>0$. The complex
function $f_j(z)$ is singular only at
\[
z=i\zeta_\pm,\quad \zeta_\pm=\frac{\pi\pm\theta}{2\sin\theta},
\]
which are $j$th-order poles, inside of $\Gamma$. By the residue theorem, we
have
\begin{equation}
\oint_\Gamma f_j(z)\,e^{ikz}\,dz =2\pi
i\left(\rho_j^{(+)}+\rho_j^{(-)}\right),  \label{e:fj}
\end{equation}
where $\rho_j^{(\pm)}$ is the residue of $f_j(z)\sin kz$ at $z=i\zeta_\pm$.
We compute
\[
\rho_2^{(\pm)} =\mp\frac{ke^{-k\zeta_{\pm}}}{4\sin^3\theta},\quad
\rho_3^{(\pm)} =\frac{ke^{-k\zeta_{\pm}}}{16\sin^5\theta}(\pm 2\cos\theta+k),
\]
so that
\[
\oint_\Gamma f_2(z)\,e^{ikz}\,dz =i\frac{\pi k}{\sin^3\theta} \exp\!\left(-%
\frac{k\pi}{2\sin\theta}\right) \sinh\!\left(\frac{k\theta}{2\sin\theta}%
\right)
\]
and
\begin{align*}
\oint_\Gamma f_3(z)\,e^{ikz}\,dz =& i\frac{\pi k}{4\sin^5\theta}
\exp\!\left(-\frac{k\pi}{2\sin\theta}\right)  \notag \\
& \times\left[ -2\sinh\!\left(\frac{k\theta}{2\sin\theta}\right)\cos\theta
+k\cosh\!\left(\frac{k\theta}{2\sin\theta}\right)\right].
\end{align*}

On the other hand, we can write
\begin{align}
\oint_\Gamma f_j(z)\,e^{ikz}\,dz =& \int_{-R}^R f_j(x)e^{ikx}dx
+i\int_0^{\pi/\sin\theta}f_j(R+iy)\,e^{ik(R+iy)}dy  \notag \\
& +\int_R^{-R} f_j(x+i\pi/\sin\theta)e^{ik(x+i\pi/\sin\theta)}dx  \notag \\
& +\int_{\pi/\sin\theta}^0 f_j(-R+iy)\,e^{ik(-R+iy)}dy.  \label{e:int3}
\end{align}
Since by $f_j(x+i\pi/\sin\theta)=f_j(x)$
\[
\int_R^{-R}f_j(x+i\pi/\sin\theta)e^{ik(x+i\pi/\sin\theta)}d x =-\exp\!\left(-%
\frac{k\pi}{\sin\theta}\right)\int_{-R}^Rf_j(x)e^{ikx}d x
\]
and the second and fourth integrals in (\ref{e:int3}) tend to zero as $%
R\rightarrow\infty$, we have
\[
\int_{-\infty}^\infty f_j(x)e^{ikx}dx =\frac{2\pi
i\left(\rho_j^{(-)}+\rho_j^{(+)}\right)} {\displaystyle 1-\exp\!\left(-\frac{%
k\pi}{\sin\theta}\right)},
\]
where we used (\ref{e:fj}). Hence, we obtain
\[
I_2=\frac{\pi k}{2\sin^3\theta}\, \mathrm{cosech}\!\left(\frac{k\pi}{%
2\sin\theta}\right) \sinh\!\left(\frac{k\theta}{2\sin\theta}\right)
\]
and
\[
I_3=\frac{\pi k}{8\sin^5\theta}\, \mathrm{cosech}\!\left(\frac{k\pi}{%
2\sin\theta}\right) \left[ -2\sinh\!\left(\frac{k\theta}{2\sin\theta}%
\right)\cos\theta +k\cosh\!\left(\frac{k\theta}{2\sin\theta}\right)\right],
\]
from which Eqs.~(\ref{e:int1}) and (\ref{e:int2}) follow.

\section{Calculation of the boundaries of the central gap via averaging}

\label{a:bound} Applying the third-order averaging method, we calculate the
boundaries of the central gap. Again, the \texttt{Mathematica} program
``haverage.m'' is used to obtain the third-order averaged systems.

\subsection{Right boundary}

Let $\omega=1-\varepsilon^2\nu$ and $\mu=\varepsilon\bar{\mu}$. Using the
transformation $(a,b)=(\varepsilon\,\xi,\varepsilon^{3/2}\eta)$, we rewrite (%
\ref{e:ab}) as
\begin{equation}
\begin{split}
\xi^{\prime}=& -2\varepsilon^{1/2}\eta +\varepsilon^{3/2}[\cos kx-\bar{\mu}%
\cos(kx+\delta)]\eta+O(\varepsilon^{5/2}), \\
\eta^{\prime}=& \varepsilon^{1/2}[\cos kx+\bar{\mu}\cos(kx+\delta)]\xi
+\varepsilon^{3/2}\left(-\nu\xi+\frac{3}{2}\xi^{3}\right)+O(%
\varepsilon^{5/2}).
\end{split}
\label{e:abr}
\end{equation}
The third-order averaged system for (\ref{e:abr}) is obtained as
\begin{equation}
\xi^{\prime}=-2\varepsilon^{1/2}\eta,\quad \eta^{\prime}=\varepsilon^{3/2}
\left[-\left(\nu-\frac{6}{k^2}(1+\bar{\mu}^2+2\bar{\mu}\cos\delta)\right)\xi
+\frac{3}{2}\xi^3\right],  \label{e:avr}
\end{equation}
where $\varepsilon^{1/2}$ was taken as the small parameter. In the averaged
system (\ref{e:avr}), the origin is a saddle if
\[
\nu>\frac{6}{k^2}(1+\bar{\mu}^2+2\bar{\mu}\cos\delta).
\]
Hence, we have the approximate right boundary of the central gap,
\[
\omega=1-\frac{6\varepsilon^2}{k^2}(1+\bar{\mu}^2+2\bar{\mu}\cos\delta).
\]

\subsection{Left boundary}

Let $\omega=-1+\varepsilon^2\nu$ and $\mu=\varepsilon\bar{\mu}$. Using the
transformation $(a,b)=(\varepsilon^{3/2}\xi,\varepsilon\eta)$, we rewrite (%
\ref{e:ab}) as
\begin{equation}
\begin{split}
\xi^{\prime}=& \varepsilon^{1/2}[\cos kx-\bar{\mu}\cos(kx+\delta)]\eta
-\varepsilon^{3/2}\left(\nu\eta+\frac{3}{2}\eta^{3}\right)+O(%
\varepsilon^{5/2}), \\
\eta^{\prime}=& -2\varepsilon^{1/2}\xi +\varepsilon^{3/2}[\cos kx+\bar{\mu}%
\cos(kx+\delta)]\xi+O(\varepsilon^{5/2}),
\end{split}
\label{e:abl}
\end{equation}
The third-order averaged system for (\ref{e:abl}) is obtained as
\[
\xi^{\prime}=\varepsilon^{3/2} \left[-\left(\nu-\frac{6}{k^2}(1+\bar{\mu}^2-2%
\bar{\mu}\cos\delta)\right)\eta -\frac{3}{2}\eta^3\right],\quad
\eta^{\prime}=-2\varepsilon^{1/2}\xi,
\]
in which the origin is a saddle if
\begin{equation}
\nu>\frac{6}{k^2}(1+\bar{\mu}^2-2\bar{\mu}\cos\delta).
\end{equation}
Thus, we have the approximate left boundary of the central gap,
\[
\omega=1-\frac{6\varepsilon^2}{k^2}(1+\bar{\mu}^2-2\bar{\mu}\cos\delta).
\]



\end{document}